\documentclass{iopart}

\usepackage{graphicx}
\usepackage{amsfonts}
\usepackage{amssymb}
\begin{document}

\title{Magnetic field induced 3D to 1D crossover in type II superconductors}

\author{T. Schneider}

\address{Physik-Institut der Universit\"{a}t Z\"{u}rich, Winterthurerstrasse 190,
CH-8057 Z\"{u}rich, Switzerland}
\ead{tschnei@physik.unizh.ch}
\begin{abstract}
We review and analyze magnetization and specific heat investigations on
type-II superconductors which uncover remarkable evidence for the magnetic
field induced finite size effect and the associated 3D to 1D crossover which
enhances thermal fluctuations. Indeed, the correlation length transverse to
the magnetic field $H_{i}$, applied along the $i$-axis, cannot grow beyond
the limiting magnetic length $L_{H_{i}}=\left( \Phi _{0}/\left(
aH_{i}\right) \right) ^{1/2}$, related to the average distance between
vortex lines. Noting that 1D is incompatible with the occurrence of a
continuous phase transition at finite temperatures, the mean-field
transition line $H_{C2}\left( T\right) $ is replaced by the 3D to 1D
crossover line $H_{p}\left( T\right) $. Since the magnetic field induced
finite size effect relies on thermal fluctuations and there enhancement
originating from the 3D to 1D crossover, its observability is not be
restricted to type-II superconductors with small correlation volume only,
including YBa$_{2}$Cu$_{4}$O$_{8}$, NdBa$_{2}$Cu$_{3}$O$_{7-\delta }$, YBa$_{2}$Cu$_{3}$O$_{7-\delta }$, and DyBa$_{2}$Cu$_{3}$O$_{7-\delta }$, where 3D-xy critical behavior was already observed in zero field. Indeed,
our analysis of the reversible magnetization of RbOs$_{2}$O$_{6}$ and the
specific heat of Nb$_{77}$Zr$_{23}$, Nb$_{3}$Sn and NbSe$_{2}$ reveals that
even in these low $T_{c}$ superconductors with comparatively large
correlation volume the 3D to 1D crossover is observable in sufficiently high
magnetic fields. Consequently, below $T_{c}$ and above $H_{pi}\left(
T\right) $ superconductivity is confined to cylinders with diameter $L_{H_{i}}$(1D) and there is no continuous phase transition in the $(H,T)$ - plane along the $H_{c2}\left( T\right) $ - line as predicted by the
mean-field treatment. Moreover we observe that the thermodynamic vortex
melting transition occurs in the 3D regime. While in YBa$_{2}$Cu$_{4}$O$_{8}$, NdBa$_{2}$Cu$_{3}$O$_{7-\delta }$, YBa$_{2}$Cu$_{3}$O$_{7-\delta }$, and DyBa$_{2}$Cu$_{3}$O$_{7-\delta }$\ it turns out to be driven by 3D-xy thermal
fluctuations, the specific heat data of the conventional type-II
superconductors Nb$_{77}$Zr$_{23}$, Nb$_{3}$Sn and NbSe$_{2}$ point to
Gaussian fluctuations. Because the vortex melting and the 3D-1D crossover
line occur at universal values of the scaling variable their relationship is
universal.
\end{abstract}

\maketitle

\begin{figure}[h]
\centering
\includegraphics[totalheight=6cm]{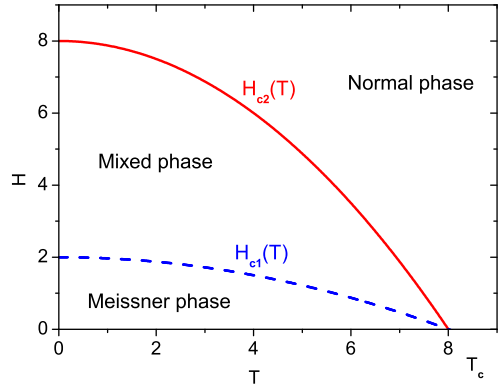}
\caption{Mean-field phase diagram comprising a normal metallic phase
at high fields and temperatures, separated by the upper
critical-field line $H_{c2}\left( T\right) $ from the mixed phase,
which in turn is separated by the lower critical-field line
$H_{c1}(T)$ from the Meissner-Ochsenfeld phase at low temperatures
and fields.} \label{fig1}
\end{figure}

The phenomenology of superconductivity is based on the
Ginzburg-Landau theory \cite{ginzburg}, which provides the starting
functional for the free energy for the charged superconductor
coupled to electromagnetism. The two fields determining the physics
of the system are the superconducting order parameter $\Psi $ and
the vector potential \textbf{A}. In its simplest version the order
parameter, $\Psi =\left\vert \Psi \right\vert \exp \left( i\varphi
\right) $, is a complex scalar. The resulting mean-field phase
diagram was constructed by Abrikosov in 1957 \cite{abrikosov} and
provided a rather accurate description for all conventional
superconductors. This mean-field $H-T$ phase diagram of a type-II
superconductor is shown schematically in Fig. \ref{fig1}.
It comprises a Meissner phase characterized by complete flux
expulsion at low magnetic fields $\left( H<H_{C1}\right) $,
separated from the mixed phase at higher fields $\left(
H>H_{c1}\right) $, where the magnetic field penetrates the
superconductor in the form of flux lines. The lower critical field
$H_{c1}\left( T\right)$ depends on the London penetration depth
$\lambda $,which is the length scale determining the electromagnetic
response of the superconductor. Since the superconducting state is a
macroscopic quantum fluid, the magnetic flux enclosed in a vortex is
quantized in units of $\Phi _{0}=hc/2e$, the flux quantum. With
increasing field the density of flux lines, forming a triangular
lattice, increases until the vortex cores overlap when the upper
critical field $ H_{c2}$ is reached. Above this field the normal
metallic phase is recovered. The upper critical field $H_{c2}$ is
determined by the correlation length of the superconductor, which is
the second fundamental length scale in the system determining which
is the characteristic distance for variations of the order
parameter.

Going beyond this traditional mean-field picture, an obvious generalization
amounts to include thermal fluctuations and disorder. Indeed, there is
considerable experimental evidence for a melting transition transforming the
vortex solid into a vortex-liquid phase \cite{blatter}. Such a melting
transition can be understood in terms of thermal fluctuations in the phase $\varphi (r)$ of the order parameter $\Psi (r)$ $=\left\vert \Psi
(r)\right\vert \exp [i\varphi (r)]$. However, in the high field regime the
phase diagram of cuprate superconductors in the $H-T$-plane is fixed by the
interplay of thermal fluctuations and disorder \cite{blatter,natter,kleinert}. Furthermore, as noted by Lee and Shenoy \cite{lee}, the fluctuations of a type-II bulk superconductor in a magnetic field become effectively one
dimensional and the superconductor behaves like an array of rods with
diameter $L_{H}\propto H^{-1/2}$ perpendicular to the applied field. As a
consequence the correlation length transverse to the applied field cannot
grow beyond the limiting magnetic length $L_{H}$ and the superconductor
undergoes with increasing field a 3D to 1D crossover. Since fluctuations
become more important with reduced dimensionality and there is the limiting
magnetic length scale $L_{H}$ it becomes clear that the upper critical field
$H_{c2}$ is an artefact of the approximations. Indeed, calculations of the
specific heat in a magnetic field which treat the interaction terms in the
Hartree approximation and extensions thereof, find that the specific heat is
smooth through the mean-field transition temperature $T_{c2}\left( H\right) $ \cite{thouless,brezin,hikami}. In the context of finite size scaling this is simply due to the fact that the correlation length of fluctuations which are
transverse to the applied magnetic field are bounded by the magnetic length $L_{H}\propto H^{-1/2}$.

In this article we review and analyze magnetization and specific heat investigations on type-II superconductors which uncover remarkable evidence for the magnetic field induced finite size effect and the associated 3D to 1D crossover which enhances the thermal fluctuations. Noting that 1D is incompatible with the occurrence of a continuous phase transition at finite temperatures, the mean-field transition line $H_{c2}\left( T\right) $ is replaced by the 3D to 1D crossover line $H_{p}\left( T\right) $. Along this line the correlation length transverse to the applied magnetic field attains the limiting magnetic length $L_{H}$ and for this reason there is no continuous phase transition. Since the magnetic field induced finite size effect relies on thermal fluctuations and there enhancement originating from
the 3D to 1D crossover, the observability of the magnetic field induced finite size effect should not be restricted to type-II superconductors with small correlation volume only. In these systems, including YBa$_{2}$Cu$_{4}$O$_{8}$ \cite{124}, NdBa$_{2}$Cu$_{3}$O$_{7-\delta}$ \cite{plackowski}, YBa$_{2}$Cu$_{3}$O$_{7-\delta}$ \cite{roulin,ts07}, and DyBa$_{2}$Cu$_{3}$O$_{7-\delta}$ \cite{garfield}, 3D-xy critical behavior was observed already in zero field. Indeed, our analysis of the reversible magnetization of RbOs$_{2}$O$_{6}$ \cite{khasanov} and the specific heat of Nb$_{77}$Zr$_{23}$ \cite{mirmelstein}, Nb$_{3}$Sn \cite{lortzNb} and NbSe$_{2}$ \cite{sanchez} reveals that even in these low $T_{c}$ superconductors with comparatively large correlation volume the 3D to 1D crossover is observable in sufficiently high magnetic fields. Moreover we observe that the thermodynamic vortex melting transition occurs in the 3D regime. While in YBa$_{2}$Cu$_{4}$O$_{8}$, NdBa$_{2}$Cu$_{3}$O$_{7-\delta }$, YBa$_{2}$Cu$_{3}$O$_{7-\delta }$, and DyBa$_{2}$Cu$_{3}$O$_{7-\delta }$\ it turns out to be driven by 3D-xy thermal fluctuations, the specific heat data of the conventional type-II superconductors Nb$_{77}$Zr$_{23}$, Nb$_{3}$Sn and NbSe$_{2}$ point to Gaussian fluctuations. Because the vortex melting and the 3D-1D crossover line occur at universal values of the scaling variable their relationship is universal.

 Next we sketch the scaling properties of the fluctuation contribution to the free energy per unit volume for a type-II superconductor in the presence of a magnetic field. On this basis we establish the equivalence with a superfluid constrained to a cylinder which unavoidably undergoes a 3D to 1D crossover. Invoking a Maxwell relation we then derive the scaling properties of the magnetization and specific heat near the 3D to 1D crossover line. We are then prepared to analyze the specific heat and reversible magnetization data of YBa$_{2}$Cu$_{4}$O$_{8}$ \cite{124}, NdBa$_{2}$Cu$_{3}$O$_{7-\delta}$ \cite{plackowski}, YBa$_{2}$Cu$_{3}$O$_{7-\delta }$ \cite{roulin,ts07}, and DyBa$_{2}$Cu$_{3}$O$_{7-\delta }$ \cite{garfield}, where critical fluctuations have been observed already in the zero field transition. Subsequently we analyze the reversible magnetization of RbOs$_{2}$O$_{6}$ \cite{khasanov} and the specific heat of Nb$_{77}$Zr$_{23}$ \cite{mirmelstein}, Nb$_{3}$Sn \cite{lortzNb} and NbSe$_{2}$ \cite{sanchez}, type-II superconductors with comparatively large correlation volume, to explore the associated enhancement of thermal fluctuations due to the 3D to 1D crossover.

It is well-known that systems of finite extent, \emph{i.e}. isolated superconducting grains, undergo a rounded and smooth phase transition \cite{fisher}. As in an infinite and homogeneous system the transition temperature $T_{c}$ is approached the correlation length $\xi $ increases strongly and diverges at $T_{c}$. However, in real systems the increase of the correlation length is limited by the spatial extent $L_{i}$ of the homogenous domains in direction $i$. In type-II superconductors, exposed to a magnetic field $H_{i}$, there is an additional limiting length scale $L_{H_{i}}=\sqrt{\Phi _{0}/\left( aH_{i}\right) }$ with $a\simeq 3.12$ \cite{bled}, related to the average distance between vortex lines \cite{bled,haussmann,lortz,parks}. Indeed, as the magnetic field increases, the density of vortex lines becomes greater, but this cannot continue indefinitely, the limit is roughly set on the proximity of vortex lines by the overlapping of their cores. Due to these limiting length scales the phase transition is rounded and occurs smoothly. Indeed, approaching $T_{c}$ from above the correlation length $\xi _{i}$ in direction $i$ increase strongly. However, due to the limiting length scales $L_{i}$ and $L_{H_{i}}=\sqrt{\Phi _{0}/\left( aH_{i}\right) }$, it is bounded and cannot grow beyond \cite{bled}
\begin{equation}
\left.
\begin{array}{c}
\xi _{i}\left( T_{p}\left( L_{i}\right) \right) =\xi _{0i}^{-}\left\vert
t_{pL}\right\vert ^{-\nu }=L_{i}, \\
\sqrt{\xi _{i}\left( T_{p}\left( H_{k}\right) \right) \xi _{j}\left(
T_{p}\left( H_{k}\right) \right) }=\sqrt{\xi _{0i}^{-}\xi _{0j}^{-}}\left\vert t_{p}\left( H_{k}\right) \right\vert ^{-\nu } \\
=\sqrt{\Phi _{0}/\left( aH_{k}\right) }=L_{H_{k}},\ i\neq j\neq k,\end{array}\right\}  \label{eq1}
\end{equation}
where
\begin{equation}
t=T/T_{c}-1.  \label{eq2}
\end{equation}
$\nu $ denotes the critical exponent of the correlation lengths $\xi _{i}$
with critical amplitude $\xi _{0i}^{-}$ below $T_{c}$.

To explore the evidence for the occurrence of these finite size effects we
concentrate on the behavior of the magnetization and the specific heat below
the zero field transition temperature $T_{c}$. Our starting point is the
fluctuation contribution to the free energy per unit volume. In the presence
of a magnetic field applied parallel to the $c$-axis it adopts the scaling
form \cite{bled,parks,book,lawrie}%
\begin{equation}
f_{s}=\frac{Q^{-}k_{B}T}{\xi _{ab}^{2}\xi _{c}}G\left( z\right) ,\: z=\frac{\xi _{ab}^{2}H_{c}}{\Phi _{0}},
\label{eq3}
\end{equation}
where $G\left( z\right) $ is a universal scaling function of its argument
and $Q^{-}$ a universal constant.  For a brief derivation of the resulting
universal relations and scaling forms for the magnetization and specific
heat we refer to the Appendix. The scaling function $G\left( z\right) $
is expected to exhibit a singularity at the universal value $z_{m}$ of the
scaling variable $z$, which describes the thermodynamic vortex melting line
\cite{lawrie}
\begin{equation}
t_{m}=\left( \frac{\left( \xi _{ab0}^{-}\right) ^{2}H_{c}}{\Phi _{0}z_{m}}
\right) ^{1/2\nu }.
\label{eq3a}
\end{equation}
There is considerable evidence that the vortex lattice melts in very clean samples via a first-order phase transition \cite{crabtree,welp,schilling}. With the introduction of random vortex pinning defects, however, the first-order melting transition becomes more continuous \cite{crabtree,safar,kwok,bouquet}, commonly referred to as the `vortex-glass' transition. There is evidence for different glassy phases (Bose glass, Bragg Glass and vortex glass) \cite{crabtree}, the detailed nature of which appears to be controversial. In any case, in sufficiently clean systems there is a thermal fluctuation driven vortex melting transition. In this case it occurs at a universal value of the scaling variable and relation (\ref{eq3a}) implies that the isotope and pressure effects on transition temperature, melting temperature and in-plane correlation length are not independent.

To recognize the implications of the magnetic field induced finite size effect, it is instructive to note that this scaling is formally equivalent to an uncharged superfluid, such as $^{4}$He, constrained to a cylinder of diameter $L_{H_{c}}=\left( \Phi _{0}/(aH_{c})\right) ^{1/2}$. Indeed, the finite size scaling theory predicts, that a system confined to a barlike geometry, $L\cdot L\cdot H$ where $H\rightarrow \infty $, an observable $O(t,L)$ scales as \cite{cardy,privman,nho}
\begin{equation}
\frac{O\left( t,L\right) }{O\left( t,\infty \right) }=f_{O}\left( y\right) ,\: y=\xi \left( t\right) /L,
\label{eq4}
\end{equation}
where $f(y)$ is the finite size scaling function. As in the confined system a 3D to 1D crossover occurs, there is a rounded transition only. Indeed, because the correlation length $\xi \left( T\right) $ cannot grow beyond $L$ there is the 3D to 1D crossover line
\begin{equation}
T_{pL}=T_{c}\left( 1-\left( \frac{\xi _{0}^{-}}{L}\right) ^{1/\nu }\right) ,
\label{eq5}
\end{equation}
which transforms with Eq. (\ref{eq1}) to
\begin{equation}
T_{p}\left( H_{c}\right) =T_{c}\left( 1-\left( \frac{\xi _{ab0}^{-}}{
L_{H_{c}}}\right) ^{1/2\nu }\right) =T_{c}\left( 1-\left( \frac{aH_{c}\left(
\xi _{ab0}^{-}\right) ^{2}}{\Phi _{0}}\right) ^{1/2\nu }\right)
\label{eq6}
\end{equation}
in the magnetic field induced case. In this context it should be recognized that 1D systems with short range interactions do not undergo a continuous phase transition at finite temperature \cite{hove2}. To uncover this crossover line we invoke Maxwell's relation $\left. \partial \left(C/T\right) /\partial H_{c}\right\vert _{T}=\left. \partial ^{2}M/\partial
T^{2}\right\vert _{H_{c}}$. Together with the scaling forms of specific heat and magnetization it yields the relation (Eq. (\ref{A.13}))
\begin{equation}
TH_{c}^{1+\alpha /2\nu }\frac{\partial \left( c/T\right) }{\partial H_{c}}=TH_{c}^{1+\alpha /2\nu }\frac{\partial ^{2}m}{\partial T^{2}}=-\frac{k_{B}A^{\pm }}{2\alpha \nu }\left\vert x\right\vert ^{1-\alpha }\frac{\partial f_{c}^{\pm }}{\partial x},
\label{eq7}
\end{equation}
where $m=M/V$ is the magnetization per unit volume. Accordingly, magnetization and specific heat data taken in different fields and plotted
as $H_{c}^{1+\alpha /2\nu }\partial ^{2}m/\partial T^{2}$ \textit{vs}. $x=t/H_{c}^{1/2\nu }$ and $H^{\left( 1+\alpha \right) /2\nu }\partial \left(C/T\right) /\partial H_{c}$ \textit{vs}. $t/H^{1/2\nu }$ should then collapse on a single curve respectively. Note that this scaling form applies as long as the limiting length is set by the magnetic field in terms of $L_{H_{c}}=\sqrt{\Phi _{0}/\left( aH_{c}\right) }$. However, in real type II superconductors the spatial extent $L_{ab}$ of the homogeneous domains is limited and by lowering the applied magnetic field $L_{H_{c}}$ approaches $L_{ab}$ unavoidably. In the regime where the growth of the in-plane correlation length is limited by $L_{H_{c}}$ these plots are expected to uncover the vortex melting transition line in terms of a peak at $x_{m}$ and the 3D to 1D crossover line in terms of a dip at $x_{p}$. This dip replaces the singularity at the so called upper critical field resulting from the Gaussian approximation and the neglect of the magnetic field induced finite size effect (see \ref{A.19}).

Another suitable observable to uncover the magnetic field induced crossover is the derivative of the specific heat with respect to temperature. According to Eq. (\ref{A.14}) it adopts the scaling form
\begin{equation}
\frac{dc}{dT}=\frac{A^{-}}{T_{c}}H_{c}^{-\left( 1+\alpha \right) /2\nu
}\left( x^{-\left( 1+\alpha \right) /2\nu }f\left( x\right) -\alpha
x^{-\alpha }\frac{df}{dx}\right) ,
\label{eq8}
\end{equation}
whereby the data $dc/dT$ for different fields should collapse on a single curve when plotted as $H_{c}^{\left( 1+\alpha \right) /2\nu }T_{c}dc/dT$ vs. $tH_{c}^{-1/2\nu }$. In the regime where the growth of the in-plane correlation length is limited by $L_{H_{c}}$ this plot should also uncover the vortex melting transition line in terms of a peak at $x_{m}$ and the 3D to 1D crossover line in terms of \ dip at $x_{p}$. This dip replaces the singularity at the so called upper critical field resulting from the
Gaussian approximation and the neglect of the magnetic field induced finite
size effect (see \ref{eqA.22}).

There is considerable evidence that cuprate superconductors with moderate anisotropy exhibit in zero field 3D-xy critical behavior \cite{plackowski,roulin,ts07,garfield,bled,lortz,parks,book,tsa,tshk,ohl,hub,kamal,jhts,mein,tsdc,tsnj}. The critical exponents are then given by \cite{kleinert3,pelissetto}
\begin{equation}
\alpha =2\nu -3\simeq -0.013,\: \nu \simeq 0.671\simeq 2/3.
\label{eq9}
\end{equation}
As disorder is concerned there is the Harris criterion \cite{harris}, which states that short-range correlated and uncorrelated disorder is irrelevant at the unperturbed critical point, provided that the specific heat exponent $\alpha $ is negative. However, when superconductivity is restricted to homogeneous domains of finite spatial extent $L_{ab,c}$, the system is inhomogeneous and the resulting rounded transition uncovers a finite size effect \cite{cardy,privman,nho} because the correlation lengths $\xi_{ab,c}=\xi _{ab0,c0}^{\pm }\left\vert t\right\vert ^{-\nu }$ cannot grow beyond $L_{ab,c}$, the respective extent of the homogenous domains. Hence, as long as $\xi _{ab,c}<L_{ab,c}$ the critical properties of the fictitious homogeneous system can be explored. There is considerable evidence that this
scenario accounts for the rounded transition seen in the specific heat \cite{book} and the magnetic penetration depths \cite{tsdc} in zero field. Nevertheless, as long as $L_{H_{k}}<L_{i}$, the magnetic field induced finite size effect sets the limiting length scale and the scaling plots $TH_{c}^{1+\alpha /2\nu }\partial ^{2}m/\partial T^{2}$ vs. $x=t/H_{c}^{1/2\nu }$ and $TH^{\left( 1+\alpha \right) /2\nu }d\left(C/T\right) /dT$ \textit{vs}. $t/H^{1/2\nu }$ should uncover the collapse on a single curve.

In classic superconductors, such as Nb$_{77}$Zr$_{23}$, Nb$_{3}$Sn and NbSe$_{2}$, the zero field specific heat measurements uncover near $T_{c}$ remarkable consistency with the standard mean-field jump \cite{mirmelstein,lortzNb,sanchez}. According to this it is unlikely that the magnetic field induced 3D to 1D crossover is associated with 3D-xy thermal fluctuations. Here we consider the contribution of Gaussian fluctuations limited by the magnetic field induced finite size effect. In this case the outlined scaling forms apply as well but the critical exponents are given by \cite{book,pelissetto}
\begin{equation}
\alpha =1/2, \: \nu =1/2.
\label{eq10}
\end{equation}

\begin{figure}[h]
\centering
\includegraphics[totalheight=6cm]{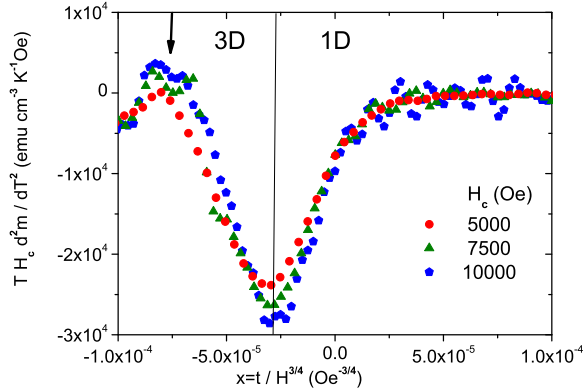}
\caption{$TH_{c}d^{2}m/dT^{2}$ \textit{vs}. $x=t/H_{c}^{3/4}$ of a YBa$_{2}$Cu$_{4}$O$_{8}$ single crystal with $T_{c}\simeq 79.6$ K taken from Weyeneth \emph{et al}. \cite{124}. The arrow marks the melting line $x_{m}\simeq -7.5\cdot 10^{-5}($ Oe$^{-3/4})$ and the vertical line $x_{p}\simeq -2.85\cdot 10^{-5}($ Oe$^{-3/4}) $ the 3D to 1D crossover line.} \label{fig2}
\end{figure}

We are now prepared to review and analyze experimental data to explore the occurrence of the magnetic field induced finite size effect and the associated 3D to 1D crossover. In Fig. \ref{fig2} we depicted the scaling plot $TH_{c}d^{2}m/dT^{2}$ \textit{vs}. $tH_{c}^{-3/4}$ of a YBa$_{2}$Cu$_{4}$O$_{8}$ single crystal taken from Weyeneth \emph{et al}.\cite{124}. It corresponds to Eq. (\ref{eq7}) with the 3D-xy exponents (Eq. (\ref{eq9})).
Apparently, the data collapses reasonably well on a single curve, consistent with dominant 3D-xy fluctuations. The occurrence of a dip and the peak, marked by the vertical line and the arrow, respectively differ drastically from the mean-field behavior where $\partial ^{2}m/\partial T^{2}=0$. Moreover, the finite depth contradicts the reputed singularity at $T_{c2}$ obtained in the Gaussian approximation \cite{prange}. The observed collapse of the data on a single curve reveals consistency with 3D-xy fluctuations down to $H_{c}=5\cdot 10^{3}$ Oe limited by the magnetic field induced finite size effect only. From $L_{H_{c}}=\left( \Phi _{0}/\left( aH_{c}\right) \right) ^{1/2}$, $a\simeq 3.12$ and $H_{c}=5\cdot 10^{3}$ Oe we obtain for the extent of the homogeneous domains the lower bound $L_{ab}>3.6\cdot 10^{-4}$ cm.
\begin{figure}[h]
\centering
\includegraphics[totalheight=6cm]{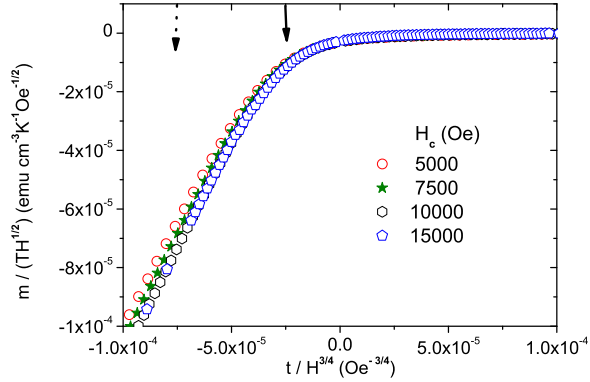}
\caption{$m/\left( TH_{c}^{1/2}\right) $ \textit{vs}. $t/H^{3/4}$ of a YBa$_{2}$Cu$_{4}$O$_{8}$ single crystal taken from \cite{124} for magnetic
fields $H_{c}$ applied along the $c$-axis. The arrows mark the melting $\left( \dot{\downarrow}\right) $ and 3D to 1D crossover line $\left(
\downarrow \right) $.} \label{fig3}
\end{figure}
In Fig. \ref{fig3}, we depicted the scaling plot $m/\left( TH_{c}^{1/2}\right)$ \textit{vs}. $t/H^{3/4}$ for YBa$_{2}$Cu$_{4}$O$_{8}$, where the magnetization data falls on a single curve (Eq. (\ref{A.1})) when 3D-xy fluctuations dominate. It is readily seen that in this plot signatures of the peak and dip structure in $TH_{c}d^{2}m/dT^{2}$ \textit{vs}. $x=t/H_{c}^{3/4}$ are hardly visible.
Though this plot reveals the dominant critical fluctuations it does not provide further insight. To relate the finite depth of the dip and the magnetic field induced finite size effect we note that at $T_{p}\left( H_{c}\right) $ the in-plane
correlation length $\xi _{ab}$ attains according to Eq. (\ref{eq1}) the limiting magnetic length $L_{H_{c}}$ so that
\begin{equation}
\xi _{ab}\left( T_{p}\right) =\xi _{ab0}^{-}\left\vert t_{p}\left(
H_{c}\right) \right\vert ^{-\nu }=L_{H_{c}}=\left( \frac{\Phi _{0}}{aH_{c}}\right) ^{1/2}.
\label{eq11}
\end{equation}
At $T_{p}$ Eq. (\ref{eq7}) yields
\begin{eqnarray}
\left. \frac{\partial ^{2}m}{\partial T^{2}}\right\vert _{T=T_{p}\left(
H_{c}\right) } &=&-\left. \frac{k_{B}A^{\pm }}{2\alpha \nu T}
H_{c}^{-1-\alpha /2\nu }\left\vert x\right\vert ^{1-\alpha }\frac{\partial
f_{c}^{\pm }}{\partial x}\right\vert _{T=T_{p}\left( H_{c}\right) }
\nonumber \\
&=&-\left. \frac{k_{B}A^{\pm }}{2\alpha \nu T}\left\vert x\right\vert
^{1-\alpha }\frac{\partial f_{c}^{\pm }}{\partial x}\left( \frac{a\xi
_{ab}^{2}\left( T\right) }{\Phi _{0}}\right) ^{1+\alpha /2\nu }\right\vert
_{T=T_{p}\left( H_{c}\right) },
\label{eq12}
\end{eqnarray}
revealing that the depth of the dip in $\partial ^{2}m(H_{c},T)/\partial
T^{2}$ \textit{vs}. $T$ is controlled by the limited growth of the
in-plane correlation length $\xi _{ab}$. This differs drastically from the
Gaussian approximation where $\partial ^{2}m(H_{c},T)/\partial T^{2}$
diverges at the so called upper critical field because the limited growth of
the correlation length is not taken into account (Appendix A). As $\xi _{ab}$
attains $T_{p}\left( H_{c}\right) $ the limiting magnetic length $L_{H_{c}}$, $\partial m(H_{c},T)/\partial T$ \textit{vs}. $T$ exhibits for fixed $H_{c}$ an inflection point and  $\partial ^{2}m(H_{c},T)/\partial T^{2}$ an extremum. Accordingly,
\begin{equation}
t_{p}\left( H_{c}\right) H_{c}^{-3/4}=-2.85\cdot 10^{-5}\left(\mathrm{Oe}^{-3/4}\right),
\label{eq13}
\end{equation}
determines the 3D to 1D crossover line along which $\xi _{ab}=L_{H_{c}}$.
Moreover, there is a peak in the 3D regime at
\begin{equation}
t_{m}\left( H_{c}\right) H_{c}^{-3/4}\simeq -8.35\cdot 10^{-5}\left(\mathrm{Oe}^{-3/4}\right),
\label{eq14}
\end{equation}
signaling the vortex melting transition (Eq. (\ref{eq3a})). Indeed, rewritten in the form, $H_{mc}\simeq 2.7\cdot 10^{5}\left( 1-T_{m}/T_{c}\right) ^{4/3}$ (Oe), it agrees reasonably well with the previous estimate $H_{mc}\simeq 1.8\cdot10^{5}\left( 1-T_{m}/T_{c}\right) ^{4/3}$ (Oe) of Katayama \textit{et al}. \cite{katayama}. Combining our estimates for the vortex melting and 3D-1D
crossover line we obtain for the universal ratios between the respective
values of the scaling variables and the reduced temperatures the estimates
\begin{equation}
\frac{z_{m}}{z_{p}}=\left( \frac{t_{p}\left( H_{c}\right) }{t_{m}\left(
H_{c}\right) }\right) ^{2\nu }\simeq 0.24,\: t_{p}\left( H_{c}\right)
/t_{m}\left( H_{c}\right) \simeq 0.34.
\label{eq15}
\end{equation}

 Another quantity, suitable to uncover the vortex melting and the 3D-1D
crossover line, is the derivative of the specific heat with respect to
temperature. It adopts the scaling form (Eq. (\ref{A.21}))
\begin{equation}
\frac{dc}{dT}=\frac{A^{-}}{T_{c}}H_{c}^{-\left( 1+\alpha \right) /2\nu
}\left( x^{-\left( 1+\alpha \right) /2\nu }f\left( x\right) -\alpha
x^{-\alpha }\frac{df}{dx}\right) .
\label{eq15a}
\end{equation}
Thus, the data for different fields should collapse on a single curve when
plotted as $H_{c}^{\left( 1+\alpha \right) /2\nu }T_{c}dc/dT$ \emph{vs}. $
tH_{c}^{-1/2\nu }$.
\begin{figure}[h]
\centering
\includegraphics[totalheight=6cm]{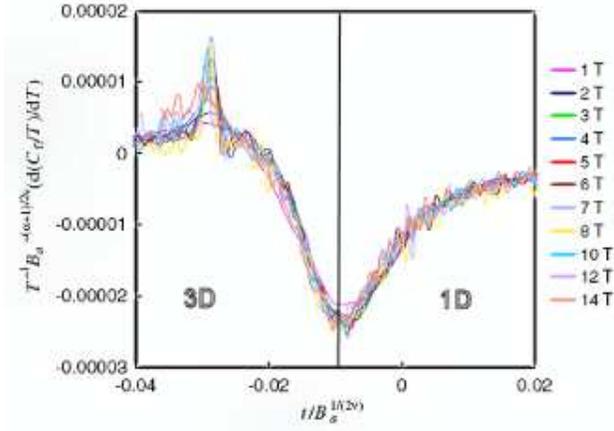}
\caption{Scaling Plot $T^{-1}H^{\left( 1+\alpha \right) /2\nu }d\left(
C/T\right) /dT$ \textit{vs}. $t/H^{1/2\nu }$ for a NdBa$_{2}$Cu$_{3}$O$_{7-\delta }$ single crystal with $T_{c}=95.5$ K taken from Plackowski\textit{et al. } \cite{plackowski}. Units are Tesla, Joule/gat and Kelvin.
The vertical line marks the 3D to 1D crossover line Eq. (\ref{eq16})) separating the 3D from the 1D regime.}
\label{fig4}
\end{figure}

In Fig. \ref{fig4} we show the data of Plackowski \textit{et al}.
\cite{plackowski} for a NdBa$_{2}$Cu$_{3}$O$_{7-\delta }$ single crystal
with $T_{c}=95.5$ K in terms of $T^{-1}H_{c}^{\left( 1+\alpha \right) /2\nu
}d\left( C/T\right) /dT$ \textit{vs}. $tH_{c}^{-1/2\nu }$ with the 3D-xy
exponents (Eq. (\ref{eq9})). Apparently, the data scales from $1$ to $14$ T
remarkably well. An essential feature is the dip reaching its minimum at
\begin{equation}
t_{p}\left( H_{c}\right) H_{c}^{-3/4}\simeq -0.0087\left( \mathrm{T}^{-3/4}\right) ,
\label{eq16}
\end{equation}
which determines the 3D to 1D crossover line. Indeed it replaces the singularity at the so called upper critical field resulting from the
Gaussian approximation and the neglect of the magnetic field induced finite size effect (see \ref{A.22}). With Eq.
(\ref{eq12}) it yields for the critical amplitude of the in-plane correlation length the estimate
\begin{equation}
\xi _{ab0}^{-}\simeq 24 \:{\mathrm{\AA}}.
\label{eq17}
\end{equation}
Furthermore, combining Eqs. (\ref{eq11}) and (\ref{eq15a}) we obtain
\begin{equation}
\left. \frac{dc}{dT}\right\vert _{T=T_{p}\left( H_{c}\right) }=\left. \frac{A^{-}}{T_{c}}\left( \frac{a\xi _{ab}^{2}\left( T\right) }{\Phi _{0}}\right)^{-\frac{1+\alpha }{2\nu }}\left( x^{-\frac{1+\alpha }{2\nu }}f\left(
x\right) -\alpha x^{-\alpha }\frac{df}{dx}\right) \right\vert
_{T=T_{p}\left( H_{c}\right) },
\label{eq18}
\end{equation}
revealing that down to $1$ T the depth of the dip is controlled by $\xi
_{ab}\left( T_{p}\left( H_{c}\right) \right) =L_{H_{c}}$, the magnetic field
induced limiting value of the in-plane correlation length. Accordingly, the
homogeneous domains exceed $L_{ab}=\left( \Phi _{0}/aH_{c}\right)
^{1/2}\simeq 2.6\cdot 10^{-6}$ cm. In addition, in the 3D regime we observe
a peak around
\begin{equation}
t_{m}\left( H_{c}\right) H_{c}^{-3/4}\simeq -0.029\left( \mathrm{T}^{-3/4}\right),
\label{eq19}
\end{equation}
which was traced back to the vortex-melting transition \cite{plackowski}.
From
\begin{equation}
z_{m}=\left( \xi _{ab0}^{-}\right) ^{2}\left\vert t_{m}\left( H_{c}\right)
\right\vert ^{-2\nu }H_{c}/\Phi _{0},\: z_{p}=\left( \xi
_{ab0}^{-}\right) ^{2}\left\vert t_{p}\left( H_{c}\right) \right\vert
^{-2\nu }H_{c}/\Phi _{0},  \label{eq20}
\end{equation}
and the estimates (\ref{eq16}) and (\ref{eq19}) the universal ratios of the
scaling variables and the reduced temperatures at the melting transition and
the 3D to 1D crossover line adopt then the values
\begin{equation}
z_{m}/z_{p}=\left( t_{p}\left( H_{c}\right) /t_{m}\left( H_{c}\right)
\right) ^{2\nu }\simeq 0.2, \:t_{p}\left( H_{c}\right) /t_{m}\left(
H_{c}\right) \simeq 0.3.
\label{eq21}
\end{equation}
With $z_{p}=1/a\simeq 0.32$ the universal value of the scaling variable, $z_{m}=$ $\left( \xi _{ab0}^{-}\right)
^{2}\left\vert t_{m}\right\vert ^{-2\nu }H_{c}/\Phi _{0}$, at
the melting transition is
\begin{equation}
z_{m}\simeq 0.06.
\label{eq22}
\end{equation}
Hence, 3D-xy fluctuations do not determine the 3D to 1D crossover line only,
but the melting line, belonging to the 3D regime, as well.
\begin{figure}[h]
\centering
\includegraphics[totalheight=6cm]{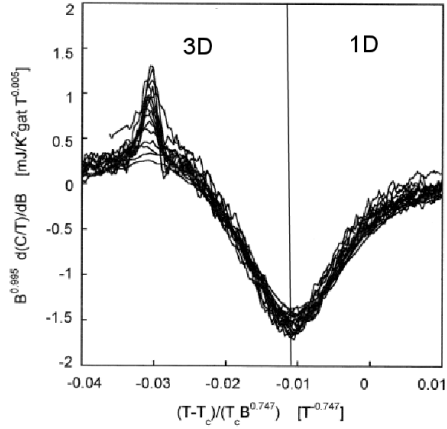}
\caption{Scaling plot of $H_{c}^{1+\alpha /2\nu }\partial \left( C/T\right)
/\partial H_{c}$ \textit{vs}. $tH_{c}^{-1/2\nu }$ for a YBa$_{2}$Cu$_{3}$O$_{6.97}$ single crystal with $T_{c}=91.1$ K and $H_{c}=1.75,$ $2.25,$ $2.75,$$3.25,$ $3.75,$ $4.25,$ $4.75,$ $5.25,$ $5.75,$ $6.25,$ $6.75,$ $7.25,$ $8,$
$9,$ $10,$ $11.25$ and $13$ T taken from Roulin \textit{et al}. \cite{roulin}. The solid line marks the 3D to 1D crossover line (Eq. (\ref{eq23})).}
\label{fig5}
\end{figure}

In Fig. 5 we depicted the scaling plot $H_{c}^{1+\alpha /2\nu }\partial\left( C/T\right) /\partial H_{c}$ \textit{vs}. $tH_{c}^{-1/2\nu }$ with 3D-xy exponents (Eq. (\ref{eq9})) for a YBa$_{2}$Cu$_{3}$O$_{6.97}$ single crystal taken from Roulin \textit{et al}. \cite{roulin}. The collapse of the data on a single curve uncovers again consistency with 3D-xy fluctuations. Moreover, we observe that down to $1.75$ T the depth of the dip is controlled by $\xi _{ab}\left( T_{p}\left( H_{c}\right) \right) =L_{H_{c}}$, the magnetic field induced limiting value of the in-plane correlation length. Accordingly, the homogeneous domains exceed $L_{ab}=\left( \Phi_{0}/aH_{c}\right) ^{1/2}\simeq 1.9\cdot 10^{-6}$ cm. The minimum of the dip at
\begin{equation}
t_{p}\left( H_{c}\right) H_{c}^{-1/2\nu }=-0.011\left( \mathrm{T}^{-3/4}\right) ,
\label{eq23}
\end{equation}
determine the 3D to 1D crossover line. With Eq. (\ref{eq11}) it yields for
the critical amplitude of the in-plane correlation length the estimate%
\begin{equation}
\xi _{ab0}^{-}\simeq 27\: {\mathrm{\AA}}.  \label{eq24}
\end{equation}
The rather sharp peak in the 3D regime at
\begin{equation}
t_{m}\left( H_{c}\right) H_{c}^{-3/4}\simeq -0.031\left(\mathrm{T}^{-3/4}\right) ,  \label{eq25}
\end{equation}
uncovers again the vortex-melting line. These estimates yield for the universal ratios the values
\begin{equation}
z_{m}/z_{p}=\left( t_{p}\left( H_{c}\right) /t_{m}\left( H_{c}\right)
\right) ^{2\nu }\simeq 0.25, \:t_{p}\left( H_{c}\right) /t_{m}\left(
H_{c}\right) \simeq 0.35,  \label{eq26}
\end{equation}
in reasonable agreement with our previous estimates for YBa$_{2}$Cu$_{4}$O$_{8}$ (Eq.(\ref{eq15})) and NdBa$_{2}$Cu$_{3}$O$_{7-\delta }$ (Eq.(\ref{eq21})). Recently it was shown that the melting line depends on pressure \cite{lortzp}. The underlying coupling between the vortex transition and the crystal lattice then demonstrates that the crystal lattice is more than a mere host for the vortices. On the other hand from the universal relation (\ref{eq3a}) it follows naturally that the relationship between $T_{c}$, $T_{m}$ and the in-plane correlation length  is fixed. From this we anticipate that if $T_{c}$ is raised (lowered) by applying uniaxial pressure, the melting line will be shifted to higher (lower) temperatures.

\begin{figure}[h]
\centering
\includegraphics[totalheight=6cm]{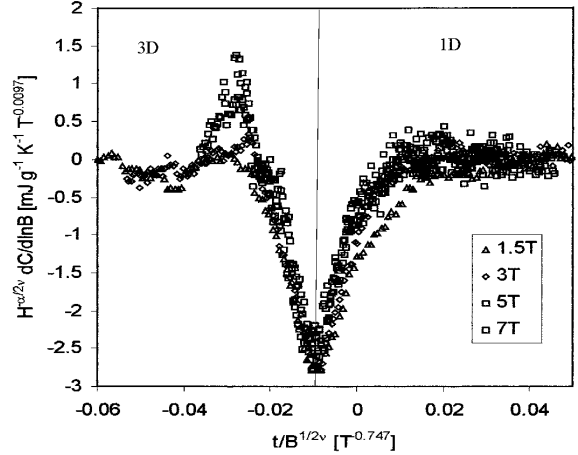}
\caption{Scaling plot $H_{c}^{1+\alpha /2\nu }\partial \left( C/T\right)
/\partial H_{c}$ \textit{vs}. $tH_{c}^{-1/2\nu }$ for a DyBa$_{2}$Cu$_{3}$O$_{6.97}$ with $T_{c}=91.05$ K taken from Garfield \textit{et al}. \cite{garfield}. The vertical line indicated the 3D to 1D crossover line (Eq. (\ref{eq27})).}
\label{fig6}
\end{figure}

In Fig. 6 we depicted the scaling plot $H_{c}^{1+\alpha /2\nu }\partial
\left( C/T\right) /\partial H_{c}$ \textit{vs}. $tH_{c}^{-1/2\nu }$ for a
for DyBa$_{2}$Cu$_{3}$O$_{7-x}$ single crystal taken from Garfield \textit{et al}. \cite{garfield}. It reveals again a reasonable collapse of the data on a single curve and with that consistency with 3D-xy fluctuations. In
addition we observe that down to $1.5$ T the depth of the dip is controlled
by $\xi _{ab}\left( T_{p}\left( H_{c}\right) \right) =L_{H_{c}}$, the
magnetic field induced limiting value of the in-plane correlation length.
Accordingly, the homogeneous domains exceed $L_{ab}=\left( \Phi
_{0}/aH_{c}\right) ^{1/2}\simeq 2.1\cdot 10^{-6}$ cm. The minimum of the dip at

\begin{equation}
t_{p}\left( H_{c}\right) H_{c}^{-3/4}=-0.0095\left(\mathrm{T}^{-3/4}\right),
\label{eq27}
\end{equation}

fixing the 3D to 1D crossover line, yields with Eq. (\ref{eq12}) for the critical amplitude of the in-plane correlation length the estimate
\begin{equation}
\xi _{ab0}^{-}\simeq 25\: {\mathrm{\AA}}.
\label{eq28}
\end{equation}
The sharp peak in the 3D regime at
\begin{equation}
t_{m}\left( H_{c}\right) H_{c}^{-3/4}\simeq -0.028\left( \mathrm{T}^{-3/4}\right),
\label{eq29}
\end{equation}
uncovers again the vortex-melting line. From the estimates (\ref{eq27}) and (\ref{eq29}) we obtain
\begin{equation}
z_{m}/z_{p}=\left( t_{p}\left( H_{c}\right) /t_{m}\left( H_{c}\right)
\right) ^{2\nu }\simeq 0.24, \:t_{p}\left( H_{c}\right) /t_{m}\left(
H_{c}\right) \simeq 0.34,
\label{eq30}
\end{equation}
in reasonable agreement with our previous estimates for the universal ratios (Eq. (\ref{eq21}) and (\ref{eq26})). Furthermore, these estimates for the 3D to 1D crossover line and the melting line agree well with $t_{p}\left(H_{c}\right) H_{c}^{-3/4}=-0.012$ (T$^{-3/4}$) and $t_{m}\left( H_{c}\right)H_{c}^{-3/4}\simeq -0.029$ (T$^{-3/4}$), derived from the respective peak positions in the total specific heat $C/T$ of an untwined YBa$_{2}$Cu$_{3}$O$_{7-x}$ single crystal with $T_{c}=91.87$ K, measured by Schilling \textit{et al}. \cite{schilling}.

Next we consider the reversible magnetization of the recently discovered superconductor RbOs$_{2}$O$_{6}$ \cite{yonezawarb}. It is a transition metal (TM) oxide which are of considerable interest because their properties range from metal-insulator transitions to colossal magnetoresistance and superconductivity. TM oxide compounds with the pyrochlore structure have long been studied and have found many applications \cite{subra}, but it is not until recently that superconductivity was discovered in one such material, namely, Cd$_{2}$Re$_{2}$O$_{7}$ at $T_{c}\simeq 1$ K \cite{hanawa,sakai,jin}. Subsequently, superconductivity was also discovered in the pyrochlore oxides KOs$_{2}$O$_{6}$ ($T_{c}\simeq 9.6$ K) \cite{yonezawak}, RbOs$_{2}$O$_{6}$ ($T_{c}\simeq 6.3$ K) \cite{yonezawarb}, and CsOs$_{2}$O$_{6}$ ($T_{c}\simeq 3.3$ K)\cite{yonezawacs}. Specific heat\cite{bruha,bruhb}, magnetization, muon-spin-rotation ($\mu $SR) studies of the magnetic penetration depth \cite{khasanova,khasanovb}, and the pressure effect measurements \cite{khasanova} on RbOs$_{2}$O$_{6}$ reveal consistent evidence for mean-field behavior, except close to $T_{c}$, where thermal fluctuations are observed. Indeed, the analysis of extended measurements of the temperature dependence of the magnetic penetration depth $\lambda $ strongly
suggest that RbOs$_{2}$O$_{6}$ falls in the universality class of charged superconductors because charged critical fluctuations were found to dominate the temperature dependence of $\lambda $ near $T_{c}$ \cite{tsrb}. It differs from the mean-field behavior observed in conventional superconductors and the uncharged (3D-xy) critical behavior found in nearly optimally doped cuprate superconductors \cite{parks,book,kamal,tsdc}, but agrees with the theoretical predictions for charged criticality \cite{kleinert2,herbut,hove}
and the charged critical behavior observed in underdoped YBa$_{2}$Cu$_{3}$O$_{6.59}$ \cite{ts697}. Noting that in the charged case the magnetic susceptibility scale as $\chi \propto t^{-\nu }$, where $\nu $ adopts the 3D-xy value \cite{kajant} (Eq.(\ref{eq9})) one expects that the scaling form
(\ref{eq7}) applies.
\begin{figure}[h]
\centering
\includegraphics[totalheight=6cm]{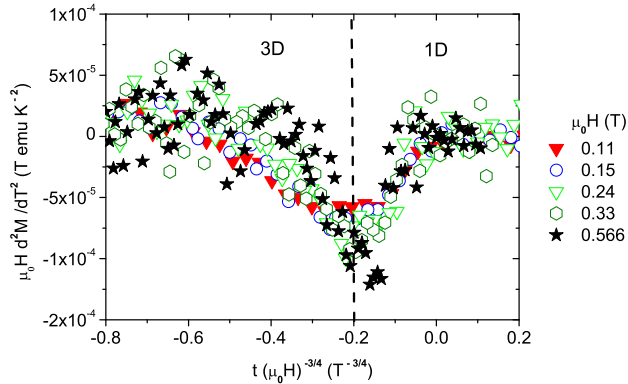}
\caption{Scaling plot $\mu _{0}Hd^{2}M/dT^{2}$ \textit{vs}. $t\left( \mu
_{0}H\right) ^{-3/4}$ for a RbOs$_{2}$O$_{6}$ powder sample with $T_{c}=6.5$
K derived from the magnetization data of Khasanov \textit{et al}. \cite{khasanov}. The vertical dashed line is the 1D to 3D crossover line (Eq. (\ref{eq31})).}
\label{fig7}
\end{figure}
In Fig. \ref{fig7} we depicted the resulting scaling plot for a RbOs$_{2}$O$_{6}$ powder sample with $T_{c}\simeq 6.5$ K. The vertical
dashed line is the 3D to 1D crossover line
\begin{equation}
t_{p}\left( \mu _{0}H\right) ^{-3/4}\simeq -0.2\left( \mathrm{T}^{-3/4}\right),
\label{eq31}
\end{equation}
yielding with Eq. (\ref{eq11}) for the critical amplitude of the correlation length the estimate
\begin{equation}
\xi _{0}^{-}\simeq 88\: {\mathrm{\AA}},
\label{eq32}
\end{equation}
in comparison with the mean-field value $\xi \left( 0\right) \simeq 74$ \AA\ \cite{bruha,bruhb}. In the 3D-xy universality class $\xi _{0}^{-}$ and $\xi_{0}^{+}$ are related by $\xi _{0}^{-}/\xi _{0}^{+}\simeq 2.21$ \cite{pelissetto}, while charged criticality exhibits inverted 3D-xy behavior \cite{inverted}. Given then the evidence for charged criticality in RbOs$_{2}$O$_{6}$ and the associated critical amplitude of the magnetic penetration depth $\lambda _{0}\simeq 1420$ \AA \cite{tsrb}, we obtain for the Ginzburg Landau parameter the estimate $\kappa \simeq 1420/\left( 2.21\cdot 88\right)
\simeq 7.3$. Accordingly, the resulting effective dimensionless charge \cite{fisherr} $\widetilde{e}=1/\kappa \simeq 0.14$ is no longer negligible and large in comparison with extreme type-II superconductors. Furthermore, according to Eqs. (\ref{eq15}), (\ref{eq21}), (\ref{eq26}), (\ref{eq30}), and (\ref{eq31}) the vortex melting line is expected to occur close to
\begin{equation}
t_{m}\left( \mu _{0}H\right) ^{-3/4}\simeq -0.63\left(\mathrm{T}^{-3/4}\right) ,
\label{eq33}
\end{equation}
where in Fig. \ref{fig7} a peak structure in the 3D regime can be anticipated. Remarkably enough, despite the comparatively large critical amplitude of the correlation length, the critical regime is accessible. Finally, from $L_{H_{c}}=\left( \Phi _{0}/\left( aH_{c}\right) \right) ^{1/2}$, $a\simeq3.12$ and $\mu _{0}H_{c}=1.1\cdot 10^{3}$ T, we obtain for the spatial extent of the homogenous domains in the RbOs$_{2}$O$_{6}$ sample the lower bound $L_{ab}>1.4\cdot 10^{-5}$ cm. In Fig. \ref{fig8}, showing $dM/dT$ \textit{vs.} $T$ for a RbOs$_{2}$O$_{6}$ powder sample at various applied magnetic fields, we observe that the aforementioned occurrence of an inflection point at $T_{p}\left( H\right)$
is well confirmed.
\begin{figure}[h]
\centering
\includegraphics[totalheight=6cm]{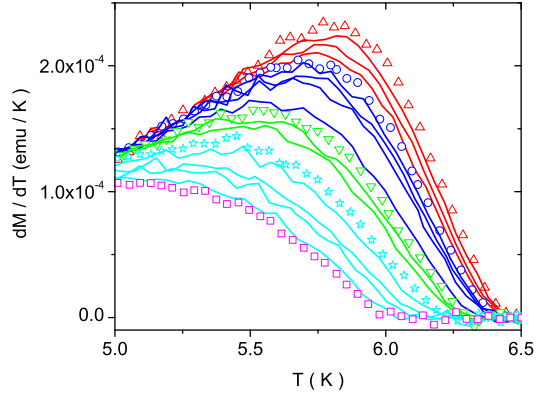}
\caption{$dM/dT$ \textit{vs.} $T$ for a RbOs$_{2}$O$_{6}$ powder sample at
various applied magnetic fields taken from Khasanov \textit{et al}. \cite{khasanov}. ($\mu _{0}H=0.11(\triangle )$, $0.12$, $0.13$, $0.14$,
$0.15(\bigcirc )$, $0.16$, $0.17$, $0.18$, $0.22$, $0.24(\triangledown )$, $0.26$, $0.28$, $0.333(\star )$, $0.366$, $0.433$, $0.466$, and $0.533$ T ($\square $)).}
\label{fig8}
\end{figure}

We have seen that in YBa$_{2}$Cu$_{4}$O$_{8}$, NdBa$_{2}$Cu$_{3}$O$_{7-\delta }$, YBa$_{2}$Cu$_{3}$O$_{6.97}$, and DyBa$_{2}$Cu$_{3}$O$_{6.97}$ where in zero magnetic field the occurrence of 3D-xy critical behavior is well established \cite{ts07,bled,lortz,parks,book,tsa,tshk,ohl,hub,kamal,jhts,mein,tsdc}, an applied magnetic field does not lead to a continuous phase transition at an upper critical field $H_{c2}$, as predicted by the mean-field approximation. Indeed, the experimental data considered here is fully consistent with a magnetic field induced finite size effect leading to a 3D to 1D crossover and a vortex melting line. Since reduced dimensionality enhances thermal fluctuations one expects that even in conventional type-II superconductors with comparatively large correlation volume the magnetic field induced finite size effect and the associated 3D to 1D crossover is at work, although the fluctuation dominated regime may not be accessible in zero magnetic field. To clarify this supposition we analyze in the following the specific heat data of Nb$_{77}$Zr$_{23}$ \cite{mirmelstein}, Nb$_{3}$Sn \cite{lortzNb}, and NbSe$_{2}$ \cite{sanchez}.

\begin{figure}[h]
\centering
\includegraphics[totalheight=6cm]{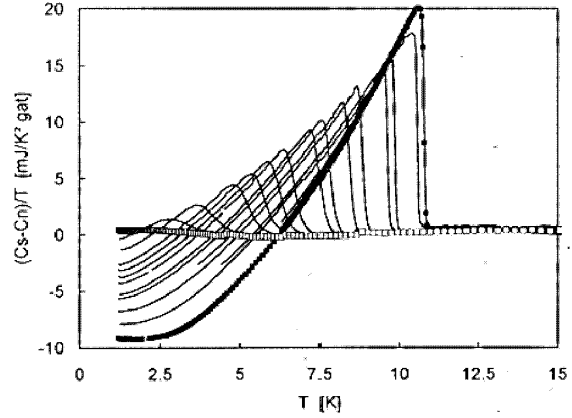}
\caption{$\Delta C/T$ \textit{vs}. $T$ for Nb$_{77}$Zr$_{23}$ taken from
Mirmelstein \textit{et al}. \cite{mirmelstein} for various applied magnetic
fields (($\blacksquare $)0, (lines) 0.2, 1, 1.2, 2, 2.4, 3, 3.3, 4, 4.4,
4.8, 5.2, 6, 6.6 T (lines), and ($\square $) 7.2 T.} \label{fig9}
\end{figure}

Nb$_{77}$Zr$_{23}$ is a cubic and isotropic type-II superconductor with $\kappa \simeq 22$. A glance to Fig. \ref{fig9} reveals that the
effect of the magnetic field on the specific heat jump is initially a shift of its position and a reduction of its height. Since the zero-field anomaly is only $50$ mK wide the fluctuation dominated regime is for sufficiently low fields not attained. Nevertheless, we observe that the jump smears out and crosses over to a peak which broadens and diminishes with increasing field.
In this view it is not surprising that the mean-field approximation
describes the data below $3$ T rather well in terms of an upper critical
field $H_{c2}$ and an assumed continuous phase transition at $T_{c}\left(
H\right) $, estimated from the "transition" midpoints in Fig. \ref{fig9} \cite{mirmelstein}. To illustrate this point and to uncover the limitations of the mean-field approach we note that the size of the specific-heat jump $\Delta C$ is related to $H_{c2}\left( T\right) $ in terms of \cite{maki}
\begin{equation}
\Delta C\left( T_{c}\left( H\right) \right) \propto \frac{T_{c}\left(
H\right) }{2\kappa ^{2}\left( T_{c}\left( H\right) \right) -1}\left. \left(
\frac{dH_{c2}}{dT}\right) ^{2}\right\vert _{T_{c}\left( H\right) },
\label{eq34}
\end{equation}
where $\kappa $ is the Ginzburg-Landau parameter.
\begin{figure}[h]
\centering
\includegraphics[totalheight=6cm]{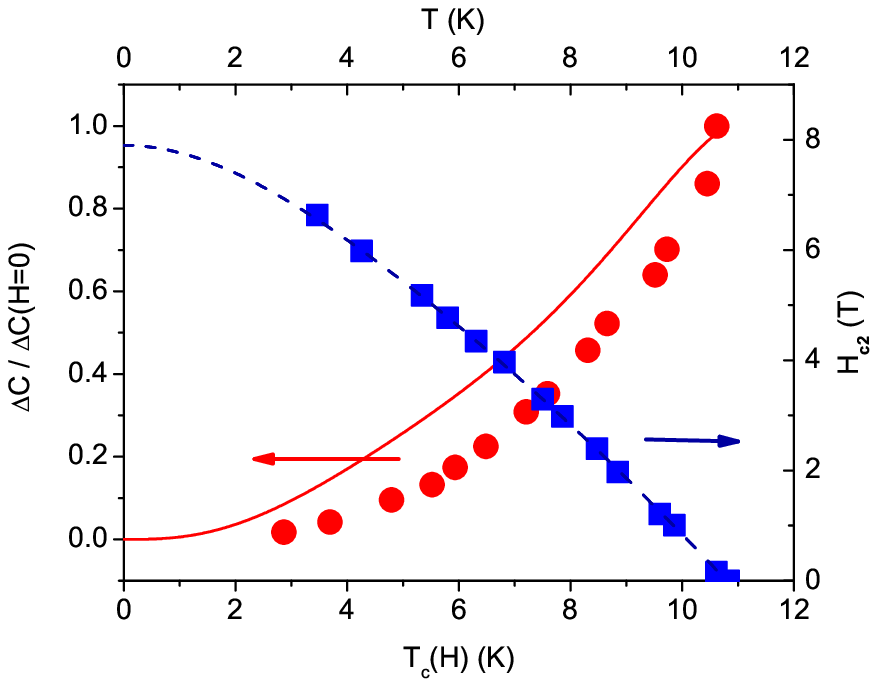}
\caption{$\Delta C/\Delta C\left( H=0\right) $ \textit{vs}. $T_{c}\left(
H\right) $ $\left( \bullet \right) $ and $H_{c2}\left( T\right) $ \textit{vs}. $T$ $\left( \blacksquare \right) $ for Nb$_{77}$Zr$_{22}$. The solid line is obtained from Eq. (\ref{eq34}) and the fitted $H_{c2}\left( T\right)
=H_{c2}\left( 0\right) \left( 1-\widetilde{t}^{2}\right) \left( 1-1.181\widetilde{t}^{2}+1.614\widetilde{t}^{3}-0.712\widetilde{t}^{4}\right) $ with $H_{c2}\left( 0\right) =7.9$T, $\widetilde{t}=T/T_{c}$, and $T_{c}=10.8$K taken from \cite{mirmelstein}, neglecting the $T$ dependence of the Ginzburg-Landau parameter $\kappa $ . The estimates for $H_{c2}\left(
T\right) $ \textit{vs}. $T$ $\left( \blacksquare \right) $ are derived from Fig. \ref{fig9} in terms of the "transition" midpoints.}
\label{fig10}
\end{figure}
In Fig. \ref{fig10} we depicted the resulting dependence $\Delta C\left( T_{c}\left( H\right) \right) /\Delta
C\left( H=0\right) $, invoking the $H_{c2}\left( T\right) $ taken from Mirmelstein \textit{et al}. \cite{mirmelstein}. Apparently, this approach describes the reduction of the jump reasonably well. Nevertheless, because an applied magnetic field drives a 3D to 1D crossover and with that enhances fluctuations, the mean-field approximation is expected to break down in sufficiently high fields. An essential signature of this scenario is the broadening of the specific heat peak, its reduction with increasing field and the field dependence of the peak location $T_{p}\left( H\right) $. From $T_{p}$ \textit{vs}. $H$ shown in Fig. \ref{fig11}a it is seen that in the regime where the jump broadens substantially ($H>2$ T) consistency with Gaussian fluctuation behavior
\begin{equation}
T_{p}(H)=11.53-1.29H,
\label{eq35}
\end{equation}
sets in. Indeed, the linear relationship corresponds to Eq.(\ref{eq6}) with $\nu =1/2$, the critical exponent of the correlation length for Gaussian fluctuations (Eq. (\ref{eq10})). This yields for the amplitude of the correlation length the estimate
\begin{equation}
\xi _{0}^{-}\simeq 86\: {\mathrm{\AA}}.
\label{eq36}
\end{equation}
\begin{figure}[h]
\includegraphics[totalheight=5cm]{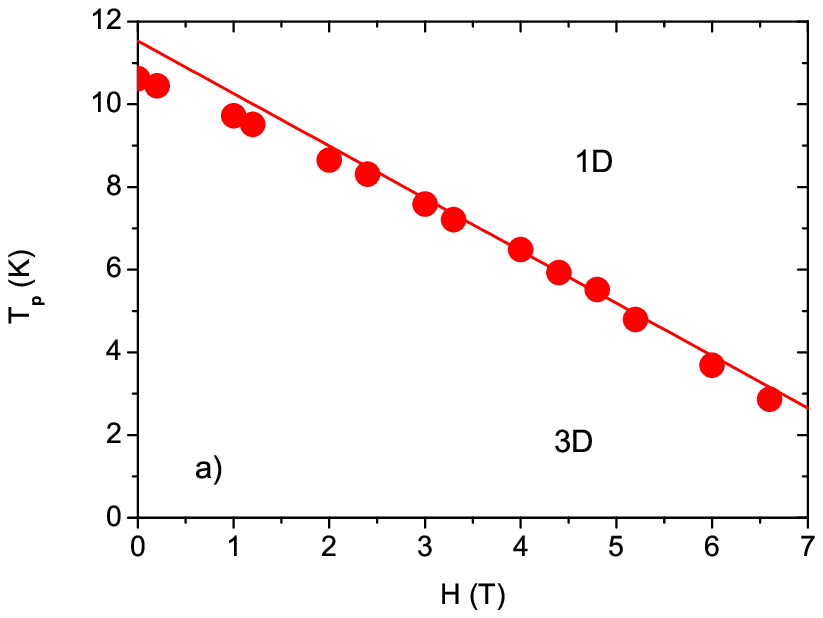}
\includegraphics[totalheight=5cm]{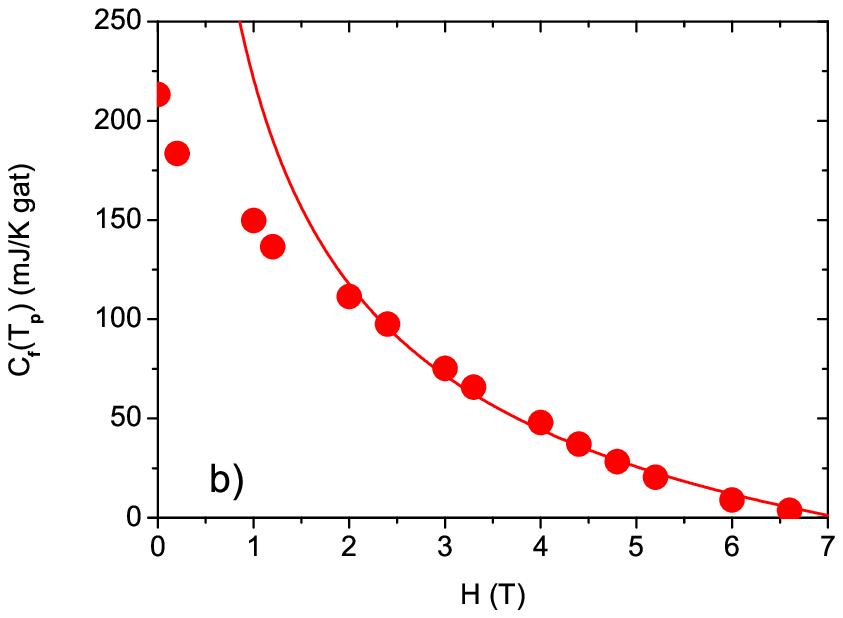}
\caption{a) 3D to 1D crossover line, $T_{p}$ \textit{vs}. $H$ for Nb$_{77}$Zr$_{22}$, derived
from the data shown in Fig. \ref{fig9}. The solid line is Eq. (\ref{eq35}); b) $C_{f}\left( T_{p}\right) $ \textit{vs}. $H$ for
Nb$_{77}$Zr$_{23}$ derived from the data shown in Fig. \ref{fig9}. The solid line is Eq. (\ref{eq39}).}
\label{fig11}
\end{figure}
In addition, Fig. \ref{fig9} reveals that the jump broadens and the height of the resulting peak decreases with increasing field. To explore the consistency of this behavior with the magnetic field induced finite size effect, implying enhanced thermal fluctuations due to the 3D to 1D crossover, we consider next the magnetic field dependence of the peak height
\begin{equation}
C_{f}\left( T_{p}\right) =C\left( T_{p}\right) -C_{b}\left( T_{p}\right) ,
\label{eq37}
\end{equation}
where $C_{b}\left( T_{p}\right) $ is the temperature dependent background. When the Gaussian scenario holds true, $C_{f}\left( T_{p}\right) $ scales
according to Eq. (\ref{A.11}) as
\begin{equation}
C_{f}\left( T_{p}\right) =\frac{A^{-}}{\alpha }\left\vert x_{p}\right\vert
^{-\alpha }f_{c}^{-}\left( x_{p}\right) H^{-\alpha /2\nu }+B,
\label{eq38}
\end{equation}
with $\alpha =1/2$ and $\nu =1/2$ characteristic for Gaussian fluctuations (Eq. (\ref{eq10})). From Fig. \ref{fig11}b, showing $C_{f}\left( T_{p}\right)$ \textit{vs}. $H$, it is seen that
\begin{equation}
C_{f}\left( T_{p}\right) =-132.44+353.64H^{-1/2},
\label{eq39}
\end{equation}
with $C_{f}\left( T_{p}\right) $ in mJ / Kgat and $H$ in T, describes the experimental data in the high field regime rather well. To substantiate the consistency with Gaussian thermal fluctuations further, we note that data taken in sufficiently high fields should collapse on a single curve when plotted as $H^{1/2}\left( C_{f}-B\right) $ \textit{vs}. $t/H$. In Fig. \ref{fig12} we observe remarkable consistency with this scaling behavior, characteristic for a magnetic field induced finite size effect associated with Gaussian fluctuations (Eq. (\ref{eq10})), enhanced by the 3D to 1D crossover. This crossovers line, $t_{p}/H=-0.11$ (T$^{-1}$), yields for the critical amplitude of the correlation length the estimate $\xi _{0}^{-}\simeq 85$ \AA, in agreement with Eq. (\ref{eq36}).
\begin{figure}[tbp]
\centering
\includegraphics[totalheight=6cm]{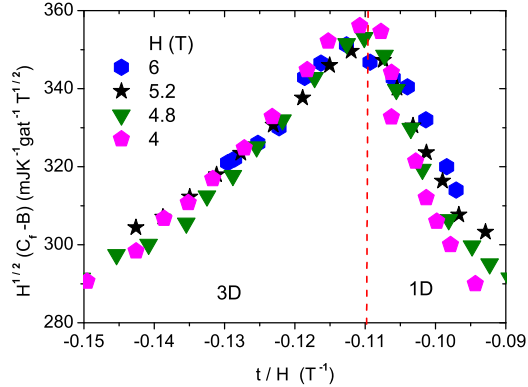}
\caption{ Scaling plot $H^{1/2}C_{f}$ \textit{vs}. $t/H$ for Nb$_{77}$Zr$_{23} $ derived from the data shown in Fig. \ref{fig9} with $B=-132.44$ (mJ / Kgat) (Eq. ( \ref{eq39})). The vertical dashed line marks $t_{p}/H=-0.11$ (T$^{-1})$, the 3D to 1D crossover line.}
\label{fig12}
\end{figure}

 While the utilization of high-$T_{c}$ cuprates made considerable progress in recent years, triniobium stannide (Nb$_{3}$Sn) is still one
of the most important materials for the practical application of superconductivity after fifty years since its discovery \cite{matthias}, most notably as superconducting wires in high-field magnets. Nb$_{3}$Sn belongs to a class of A$_{3}$B binary intermetallic compounds with the A15 or $\beta $ -tungsten structure, where those based on Nb give rise to a subclass that includes Nb$_{3}$Ge$_{2}$ \cite{matthias2}, having the highest $T_{c}$ ($\simeq 23$ K) for more than thirty years, and therefore it was once a subject of intensive study until '70s \cite{testardi}. Surprisingly enough, it seems that the fundamental superconducting properties of Nb$_{3}$Sn are
not fully understood.
\begin{figure}[tbp]
\centering
\includegraphics[totalheight=6cm]{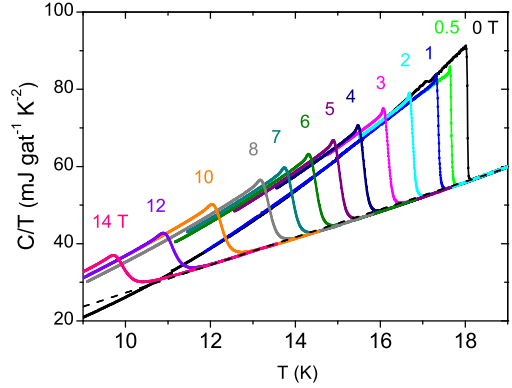}
\caption{ Total specific heat of a single crystal of Nb$_{3}$Sn in
fields from $0-14$ T measured by an AC technique taken from Lortz
\textit{et al}. \cite{lortzNb}. The dashed line is
$C_{b}/T=-8.72+3.6T$ (mJ gat$^{-1}$K$^{-2}$), taken as temperature dependent background.}
\label{fig13}
\end{figure}
In Fig. \ref{fig13} we depicted the total specific heat of a Nb$_{3}$Sn single crystal measured by Lortz \emph{et al}. \cite{lortzNb} in fields from $0-14$ T. In zero field consistency with the typical mean-field jump is
observed and there appears to be no signature of fluctuations as the superconducting phase is entered from the normal state. However, with
increasing magnetic field the jump broadens and the height of the resulting peak decreases. To explore the consistency of this behavior with the magnetic field induced finite size effect, we explore the scaling of the peak position $T_{p}(H)$ and of the fluctuation contribution $C_{f}\left(T_{p}\right) $ to the specific heat.
\begin{figure}[h]
\includegraphics[totalheight=5cm]{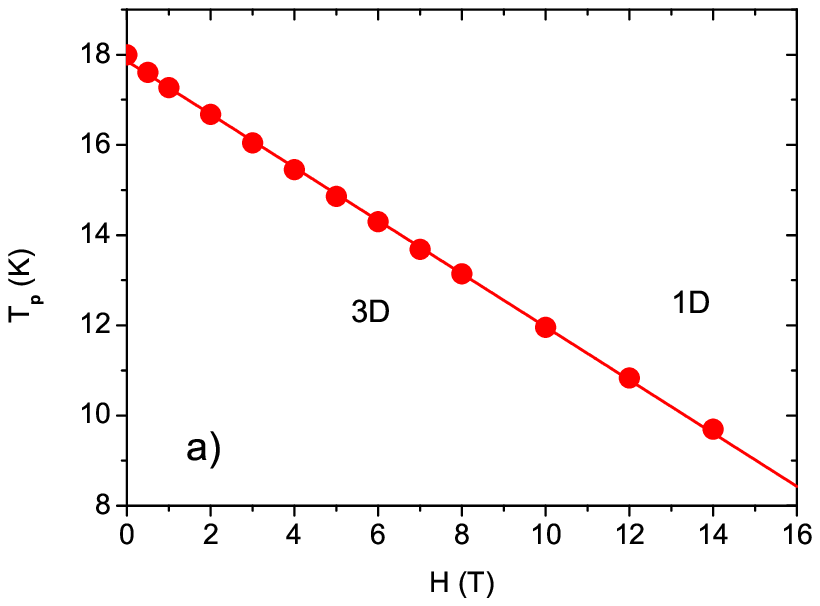}
\includegraphics[totalheight=5cm]{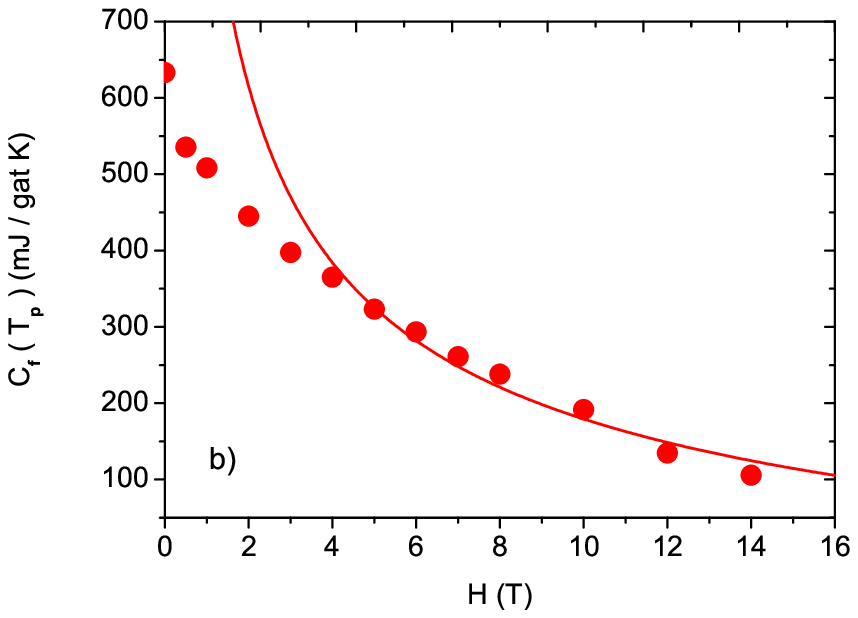}
\caption{ a) 3D to 1D crossover line, $T_{p}$ \textit{vs}. $H$, for Nb$_{3}$Sn deduced from the
specific heat data shown in Fig. \ref{fig13}. The solid line is Eq. (\ref{eq40}); b) $C_{f}\left( T_{p}\right) $ \textit{vs}. $H$ for
Nb$_{3}$Sn derived from the data shown in Fig. \ref{fig13}.The solid
line is Eq. (\ref{eq42}) with $C_{b}\left( T_{p}\right) /T_{p}=-8.72+3.6T_{p}$ (mJ gat$^{-1}$K$^{-2}
$) indicated in Fig. \ref{fig13}.}
\label{fig14}
\end{figure}
From Fig. \ref{fig14}a it is seen that the linear relationship
\begin{equation}
T_{p}\left( H\right) =17.86\left( 1-0.033H\right) \left( \mathrm{K}\right),\: t_{p}/H=-0.033\left( \mathrm{T}^{-1}\right),
\label{eq40}
\end{equation}
for the 3D to 1D crossover line describes the data rather well. This uncovers with Eq. (\ref{eq6}) Gaussian fluctuations (Eq. (\ref{eq10})) with
\begin{equation}
\xi _{0}^{-}\simeq 46 \:{\mathrm{\AA}},
\label{eq41}
\end{equation}
for the critical amplitude of the correlation length. When this Gaussian scenario holds true, $C_{f}\left( T_{p}\right)$
should then scale according to Eq. (\ref{eq10}). From Fig. \ref{fig14}b, showing $C_{f}\left( T_{p}\right) $ \textit{vs}. $H$, it is seen
that
\begin{equation}
C_{f}\left( T_{p}\right) =-173.3+1115H^{-1/2}\: (\mathrm{mJ/Kgat}),
\label{eq42}
\end{equation}
describes the experimental data for sufficiently high fields rather well and confirms with that the magnetic field induced finite size effect scenario driven by Gaussian fluctuations.
\begin{figure}[h]
\centering
\includegraphics[totalheight=6cm]{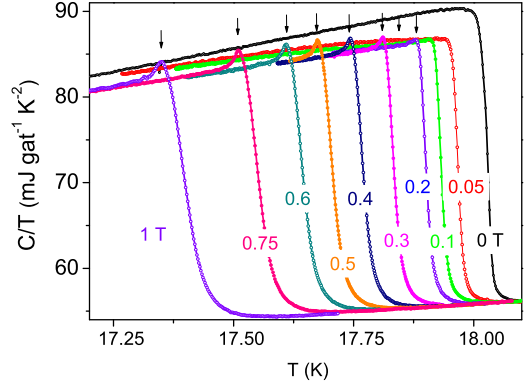}
\caption{ Total specific heat of a single crystal of Nb$_{3}$Sn in
fields from $0$ to $1$T taken from Lortz \textit{et
al}. \cite{lortzNb}. The arrows mark the location of the vortex
melting transition.} \label{fig15}
\end{figure}

\begin{figure}[h]
\centering
\includegraphics[totalheight=6cm]{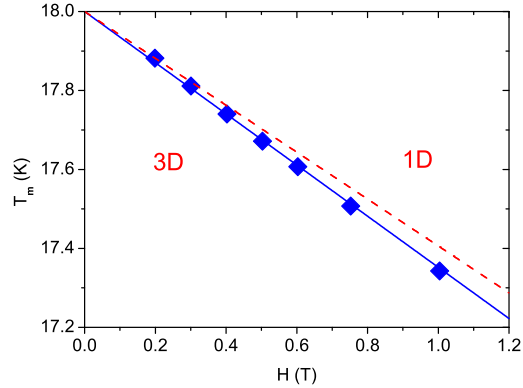}
\caption{ Vortex melting line, $T_{m}$ \textit{vs}. $H$ of Nb$_{3}$Sn derived from the location of the marked anomalies in Fig.
\ref{fig15}. The solid line is the melting line (Eq. (\ref{eq43})) and the dashed one the 3D to 1D crossover line (Eq. (\ref{eq40})).}
\label{fig16}
\end{figure}

Furthermore, the more detailed plot of the specific heat coefficient depicted in Fig. \ref{fig15} uncovers in the low field regime and below the mean-field jump a weak peak. This anomaly, marked by arrows, was traced back to the vortex melting transition \cite{lortzNb} and confirms with that the enhancement of thermal fluctuations with increasing field strength. The resulting melting line, consistent with
\begin{equation}
T_{m}\left( H\right) =18\left( 1-0.036H\right) ,\: t_{m}/H=-0.036 \left( \mathrm{T}^{-1}\right),
\label{eq43}
\end{equation}
is shown in Fig. \ref{fig16} and reveals again consistency with Gaussian fluctuations. For comparison we included the 3D to 1D
crossover line, disclosing that the vortex melting transition occurs in the 3D regime. From Eq. (\ref{eq40}) and (\ref{eq43}) we obtain in the Gaussian case with $z_{m}/z_{p}=\left( t_{p}/t_{m}\right) ^{2\nu }$ (Eq. (\ref{eq30})) for the universal ratio the estimate
\begin{equation}
t_{p}\left( H\right) /t_{m}\left( H\right) =z_{m}/z_{p}\simeq 0.92,
\label{eq44}
\end{equation}
which differs substantially from the 3D-xy value $t_{p}\left(H_{c}\right) /t_{m}\left( H_{c}\right) \simeq 0.32$ (Eqs. (\ref{eq21}), (\ref{eq26}) and (\ref{eq30})).

\begin{figure}[h]
\centering
\includegraphics[totalheight=6cm]{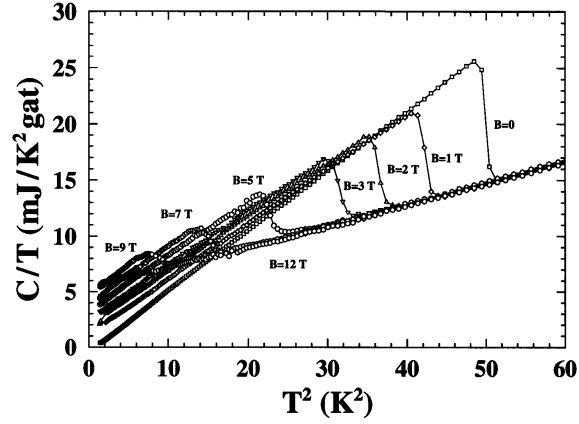}
\caption{ Specific heat of a NbSe$_{2}$ single crystal in various
magnetic fields applied parallel to the layers taken from Sanchez
\textit{et al}.\cite{sanchez}}
\label{fig17}
\end{figure}

 So far we considered either anisotropic superconductors in magnetic fields applied along the $c$-axis or isotropic materials. A suitable
anisotropic type-II superconductor where the effects of the magnetic field on the specific heat, for fields applied parallel and
perpendicular to the layers, have been studied in some detail is NbSe$_{2}$ \cite{sanchez}. Moreover, the relevance of fluctuations
was established already some time ago by Frindt \cite{frindt} in terms of the dependence of $T_{c}$ on the crystal thickness, reduced
below six NbSe$_{2}$ layers. In Fig. \ref{fig17} we reproduced the specific heat data of Sanchez \textit{et al}. \cite{sanchez} for a
NbSe$_{2}$ single crystal in various magnetic fields applied parallel to the layers.
In analogy to Nb$_{77}$Zr$_{23}$ (Fig. \ref{fig9}) and Nb$_{3}$Sn (Fig. \ref{fig13}) the jump broadens and its height decreases with
increasing magnetic field. The same behavior was observed for fields applied perpendicular to the layers \cite {sanchez}.
\begin{figure}[h]
\centering
\includegraphics[totalheight=6cm]{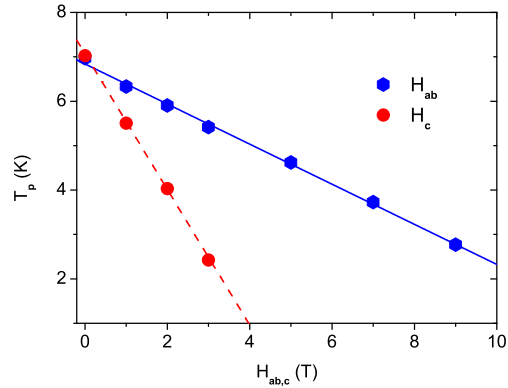}
\caption{ 3D to 1D crossover lines, $T_{p}$ \textit{vs}. $H_{ab,c}$, for NbSe$_{2}$ deduced
from the specific heat data of Sanchez \textit{et al}. \cite{sanchez}. The solid line is Eq. (\ref{eq45}), the 3D to 1D
crossover line for fields applied parallel to the $ab$-plane and the
dashed one is Eq. (\ref{eq46}) the dimensional crossover line for
fields applied parallel to the $c$-axis.}
\label{fig18}
\end{figure}
In Fig. \ref{fig18} we depicted $T_{p}$ vs. $H_{ab}$ and $H_{c}$ for NbSe$_{2}$ derived from Fig. \ref{fig17} and the data of Sanchez \textit{et al}. \cite{sanchez}. The solid and dashed lines are
\begin{eqnarray}
T_{p}\left( H_{ab}\right) &=&6.84\left( 1-0.066H_{ab}\right)\: (\mathrm{K}),
\label{eq45}
\\
T_{p}\left( H_{c}\right) &=&7.07\left( 1-0.216H_{ab}\right)\: (\mathrm{K}),
\label{eq46}
\end{eqnarray}
revealing consistency with a 3D to 1D crossover line associated with Gaussian fluctuations (Eq. (\ref{eq10})). Invoking Eq. (\ref{eq1})
we obtain for the critical amplitudes of the correlation lengths and the anisotropy $\gamma $ the estimates
\begin{equation}
\xi _{ab0}^{-}\simeq 120 \:{\mathrm{\AA}},\: \left( \xi _{ab0}^{-}\xi
_{c0}^{-}\right) ^{1/2}=66 \:{\mathrm{\AA}},  \label{eq47}
\end{equation}
and
\begin{equation}
\xi _{c0}^{-}\simeq 36 \:{\mathrm{\AA}},\: \gamma =\xi _{ab0}^{-}/\xi
_{c0}^{-}\simeq 1.8,
\label{eq48}
\end{equation}
compared to the previous estimate, $\gamma \simeq 2.4$ \cite{soto}, for the ansisotropy.
When this crossover line stems from Gaussian fluctuations the extension of Eq. (\ref{A.11}) to anisotropic systems implies that
the peak height $C_{f}\left( T_{p}\left( H_{ab,c}\right) \right) $ scale as
\begin{equation}
c_{f}\left( T_{p}\left( H_{c}\right) \right) =\frac{A^{-}}{\alpha }
f_{c}^{-}\left( x_{pc}\right) \left( \frac{\left( \xi
_{ab0}^{-}\right) ^{2}aH_{c}}{\Phi _{0}}\right) ^{-\alpha /2\nu
}+B_{c},
\label{eq49}
\end{equation}
for fields applied parallel to the $c$-axis where
\begin{equation}
x_{pc}=\frac{t_{p}}{H_{c}^{1/2\nu }}=\left( \frac{\left( \xi
_{ab0}^{-}\right) ^{2}a}{\Phi _{0}}\right) ^{1/2\nu } \label{eq50}
\end{equation}
and for fields applied in the $ab$-plane
\begin{equation}
c_{f}\left( T_{_{p}}\left( H_{ab}\right) \right)
=\frac{A^{-}}{\alpha } f_{c}^{-}\left( x_{pab}\right) \left(
\frac{\xi _{ab0}^{-}\xi _{c0}^{-}aH_{ab}}{\Phi _{0}}\right)
^{-\alpha /2\nu }+B_{ab},
\label{eq51}
\end{equation}
where
\begin{equation}
x_{pab}=\frac{t_{p}}{H_{c}^{1/2\nu }}=\left( \frac{\xi _{ab0}^{-}\xi
_{c0}^{-}a}{\Phi _{0}}\right) ^{1/2\nu },
\label{eq52}
\end{equation}
with $\alpha =1/2$ and $\nu =1/2$ for Gaussian fluctuations (Eq. (\ref{eq10})). In Fig. \ref{fig19} we show $C_{f}\left( T_{p}\left(
H_{ab,c}\right) \right) $ \textit{vs}. $ H_{ab,c}$ derived from the data of Sanchez \textit{et al}. \cite{sanchez}. For comparison we
included the fits
\begin{eqnarray}
C_{f}\left( T_{p}\left( H_{ab}\right) \right)
&=&-9.13+66.45H_{ab}^{-1/2}\: (\mathrm{mJ/K gat}),
\label{eq53} \\
C_{f}\left( T_{p}\left( H_{c}\right) \right)
&=&-34.97+66.8H_{c}^{-1/2}\: (\mathrm{mJ/K gat}),
\label{eq54}
\end{eqnarray}
which describe the sparse data reasonably well and confirm with that the 3D to 1D crossover arising from the magnetic field induced
finite size effect tuned by Gaussian fluctuations. In addition, the ratio $66.45/66.8\simeq \xi _{ab0}^{-}/\xi _{c0}^{-}\simeq \gamma
\simeq 1$, provides an independent estimate of the anisotropy.
\begin{figure}[h]
\centering
\includegraphics[totalheight=6cm]{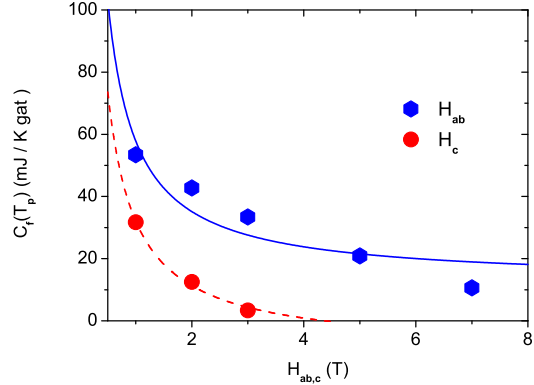}
\caption{$C_{f}\left( T_{p}\right) $ \textit{vs}. $H_{ab,c}$ for NbSe$_{2}$
derived from the specific heat data of Sanchez \textit{et} \textit{al}.\cite{sanchez}. The solid line is Eq. (\ref{eq53}) with $C_{b}\left( T_{p}\left(H_{ab}\right) \right) /T_{p}=4.95+0.196T_{p}^{2}$ (mJ/ K$^{2}$gat) and the dashed one Eq. (\ref{eq54}) with $C_{b}\left( T_{p}\left( H_{c}\right) \right) /T_{p}\left(H_{c}\right) =5.06+0.193T_{p}^{2}$ (mJ/ K$^{2}$gat) taken as temperature dependent background.}
\label{fig19}
\end{figure}
While the rather limited resolution of the specific heat data of NbSe$_{2}$ (Fig. \ref{fig17}) does not reveal any signature of the thermodynamic vortex melting transition, mode locking and dc transport measurements uncovered this transition over a wide magnetic field and temperature range \cite{kokubo}. Accordingly, the universal ratio between the 3D-1D crossover and vortex melting line (Eq. (\ref{eq44})) is expected to apply.

 The characteristic features of the magnetic field induced finite size effect on the specific heat, broadening of the peak and the
reduction of its height with increasing field, have also been observed in ZrB$_{12}$ \cite{wang} and $\kappa $
-(BEDT-TTF)$_{2}$Cu(NCS)$_{2}$ \cite{lortzor}. In contrast to the systems considered so far, these materials exhibit a richer
response to an applied magnetic field. While in most type-II superconductors the orbital-pair breaking mechanism dominates, this
is no longer the case in $\kappa $ -(BEDT-TTF)$_{2}$Cu(NCS)$_{2}$. In parallel fields above $21$ T the Pauli-paramagnetic pair-breaking
effects appear and there is evidence for the Fulde-Ferrel-Larkin-Ovchinnikov phase \cite{fulde,larkin}.
Furthermore, clean ZrB$_{12}$ with $T_{c}\simeq 6$ K is a superconductor whose Ginzburg-Landau parameter $\kappa \simeq 0.65$
is close to the border value $2^{-1/2}\simeq 0.7$. Since $\kappa $ varies with temperature the material crosses over from type-I to
type-II as the temperature is lowered.

 We have seen that measurements of thermodynamic properties have evolved into a major experimental tool in the quest for an understanding of
type-II superconductors. Noting that the influence of magnetic fields on the superconducting state is of high technological relevance this issue is not of fundamental interest only. Nevertheless, the effects of thermal fluctuations were not fully identified over several decades. Particularly in classical superconductors, including Nb$_{77}$Zr$_{23}$, Nb$_{3}$Sn and NbSe$_{2}$, interpretations based on the standard mean-field approximation were considered to be satisfactory. By contrast, our analysis of the specific heat data uncovered in sufficiently high fields remarkable consistency with the magnetic field induced finite size effect, giving rise to a 3D to 1D
crossover which enhances thermal fluctuations. While in YBa$_{2}$Cu$_{4}$O$_{8}$, NdBa$_{2}$Cu$_{3}$O$_{7-\delta }$, YBa$_{2}$Cu$_{3}$O$_{7-\delta }$, and DyBa$_{2}$Cu$_{3}$O$_{7-\delta }$\ 3D-xy-, in RbOs$_{2}$O$_{6}$ inverted
3D-xy-thermal fluctuations were shown to drive this crossover, the specific heat data of the conventional type-II superconductors Nb$_{77}$Zr$_{23}$, Nb$_{3}$Sn and NbSe$_{2}$ point to Gaussian fluctuations. In any case, whenever this crossover occurs there is no continuous phase transition in the $H-T$ -plane along a $H_{c2}\left( T\right) $-line, as predicted by the mean-field treatment. Instead there is the 3D to 1D crossover line $H_{pi}\left(T\right) =\left( \Phi _{0}/\left( a\xi _{j0}^{-}\xi _{k0}^{-}\right) \right)
(1-T/T_{c})^{1/2\nu }$ with $i\neq j\neq k$ and $\xi _{i0,j0,k0}^{-}$ denoting the critical amplitude of the correlation length below $T_{c}$ with critical exponent $\nu $. Accordingly, below $T_{c}$ and above $H_{pi}\left(T\right) $ superconductivity is confined to cylinders with diameter $L_{H_{i}}\propto H_{i}^{-1/2}$ whereupon the system becomes 1D. However, 1D systems with short range interactions do not undergo a continuous phase transition at finite temperature \cite{hove} and with that there is no continuous phase transition in the $H-T$-plane above the 3D to 1D crossover line. Furthermore, we have shown that the thermodynamic vortex melting
transition occurs in the 3D regime. While in YBa$_{2}$Cu$_{4}$O$_{8}$, NdBa$_{2}$Cu$_{3}$O$_{7-\delta }$, YBa$_{2}$Cu$_{3}$O$_{7-\delta }$, and DyBa$_{2}$Cu$_{3}$O$_{7-\delta }$\ it is driven by 3D-xy thermal fluctuations, the
specific heat data of the conventional type-II superconductors Nb$_{77}$Zr$_{23}$, Nb$_{3}$Sn and NbSe$_{2}$ point to Gaussian fluctuations. In any case, because the vortex melting transition and the 3D-1D crossover occur at universal values of the scaling variable $z$, the ratio $z_{m}/z_{p}=\left(t_{p}\left( H_{c}\right) /t_{m}\left( H_{c}\right) \right) ^{2\nu }$ should
be universal as well. Our analysis revealed, $z_{m}/z_{p}=\left( t_{p}\left(H_{c}\right) /t_{m}\left( H_{c}\right) \right) ^{2\nu }\simeq 0.23$ (Eqs. (\ref{eq21}), (\ref{eq26}) and (\ref{eq30})), when 3D-xy fluctuations dominate and $z_{m}/z_{p}=t_{p}\left( H\right) /t_{m}\left( H\right) \simeq 0.92$ (Eq. (\ref{eq44})) in the Gaussian case. Thus it appears that thermal fluctuations, enhanced by the 3D to 1D crossover,  are important not only in high-temperature superconductors but also in conventional type-II superconductors with comparatively large correlation volume. This observation opens up a window onto the universal properties mediated by thermal fluctuations. and allows to probe the thermodynamic relevant spatial extent of the homogenous domains. From this perspective, more extended high resolution specific heat and reversible magnetization measurements on type-II superconductors and their analyzes along the lines outlined here will certainly be necessary to unravel the details of the universal properties associated with the 3D to 1D crossover and the vortex melting transition in type II superconductors.

\ack
Over the years, we have benefited from numerous scientific discussions on
this topic with H. Keller, R. Khasanov, R. Lortz, C. Meingast, K. A.
M\"{u}ller, J. Roos, J. M. Singer, and S. Weyeneth. Moreover, the
author is grateful to R. Khasanov, and R. Lortz for
providing experimental data.
\appendix
\setcounter{section}{1}
\section*{Appendix A}
When in a type II superconductor thermal fluctuations dominate and the
coupling to the charge is negligible the free energy per unit volume adopts
in the presence of a magnetic field $H_{c}$ applied parallel to the $c$-axis
the scaling form (\ref{eq3}). This leads for the magnetization per unit
volume, $m=M/V=-\partial f_{s}/\partial H_{c}$ , the scaling expression \cite{ts07,parks,book,jhts}
\begin{eqnarray}
\frac{m}{TH^{1/2}} &=&-\frac{Q^{\pm }k_{B}\xi _{ab}}{\Phi _{0}^{3/2}\xi _{c}}F^{\pm }\left( z\right) , F^{\pm }\left( z\right) =z^{-1/2}\frac{dG^{\pm }}{dz},  \nonumber \\
z &=&x^{-1/2\nu }=\frac{\left( \xi _{ab0}^{\pm }\right) ^{2}\left\vert
t\right\vert ^{-2\nu }H_{c}}{\Phi _{0}}.
\label{A.1}
\end{eqnarray}
$Q^{\pm }$ is a universal constant and $G^{\pm }\left( z\right) $ a universal scaling function of its argument, with $G^{\pm }\left( z=0\right)=1$. $\gamma =\xi _{ab}/\xi _{c}$ denotes the anisotropy, $\xi _{ab}$ the zero-field in-plane correlation length and $H_{c}$ the magnetic field applied along the $c$-axis. In terms of the variable $x$ the scaling form (\ref{eq1}) is similar to Prange's \cite{prange} result for Gaussian fluctuations. Approaching $T_{c}$ the in-plane correlation length diverges as
\begin{equation}
\xi _{ab}=\xi _{ab0}^{\pm }\left\vert t\right\vert ^{-\nu },\:
t=T/T_{c}-1, \:\pm =sgn(t).  \label{A.2}
\end{equation}
Supposing that 3D-xy fluctuations dominate the critical exponents are given by \cite{kleinert3,pelissetto}
\begin{equation}
\nu \simeq 0.671\simeq 2/3,\: \alpha =2\nu -3\simeq -0.013,
\label{A.3}
\end{equation}
and there are the universal critical amplitude relations \cite{ts07,parks,book,jhts,kleinert3,pelissetto}
\begin{equation}
\frac{\xi _{ab0}^{-}}{\xi _{ab0}^{+}}=\frac{\xi _{c0}^{-}}{\xi _{c0}^{+}}\simeq 2.21,\: \frac{Q^{-}}{Q^{+}}\simeq 11.5,
\frac{A^{+}}{A^{-}}=1.07,
\label{A.4}
\end{equation}
and
\begin{eqnarray}
A^{-}\xi _{a0}^{-}\xi _{b0}^{-}\xi _{c0}^{-} &\simeq &A^{-}\left( \xi
_{ab0}^{-}\right) ^{2}\xi _{c0}^{-}=\frac{A^{-}\left( \xi _{ab0}^{-}\right)
^{3}}{\gamma }  \nonumber \\
&=&\left( R^{-}\right) ^{3},\: R^{-}\simeq 0.815,  \label{A.5}
\end{eqnarray}
where $A^{\pm }$ is the critical amplitude of the specific heat singularity, defined as
\begin{equation}
c=\frac{C}{Vk_{B}}=\frac{A^{\pm }}{\alpha }\left\vert t\right\vert ^{-\alpha
}+B,  \label{A.6}
\end{equation}
where $B$ denotes the background. Furthermore, in the 3D-xy universality class $T_{c}$, $\xi _{c0}^{-}$ and the critical amplitude of the in-plane penetration depth $\lambda _{ab0}$ are not independent but related by the universal relation \cite{ts07,parks,book,jhts,pelissetto},
\begin{equation}
k_{B}T_{c}=\frac{\Phi _{0}^{2}}{16\pi ^{3}}\frac{\xi _{c0}^{-}}{\lambda
_{ab0}^{2}}=\frac{\Phi _{0}^{2}}{16\pi ^{3}}\frac{\xi _{ab0}^{-}}{\gamma
\lambda _{ab0}^{2}}.
\label{A.7}
\end{equation}
From this universal relation it follows naturally that the isotope and pressure effects on the transition temperature, the correlation lengths, the anisotropy and the magnetic penetration depths are not independent \cite{tsiso,124press}. Furthermore, the existence of the magnetization at $T_{c}$, of the penetration depth below $T_{c}$ and of the magnetic susceptibility above $T_{c}$ imply the following asymptotic forms of the scaling function \cite{ts07,parks,book,jhts}
\begin{eqnarray}
Q^{\pm }\left. \frac{1}{\sqrt{z}}\frac{dG^{\pm }}{dz}\right\vert
_{z\rightarrow \infty } &=&Q^{\pm }c_{\infty }^{\pm },  \nonumber \\
Q^{-}\left. \frac{dG^{-}}{dz}\right\vert _{z\rightarrow 0}
&=&Q^{-}c_{0}^{-}\left( \ln z+c_{1}\right) ,  \nonumber \\
Q^{+}\left. \frac{1}{z}\frac{dG^{+}}{dz}\right\vert _{z\rightarrow 0}
&=&Q^{+}c_{0}^{+},
\label{A.8}
\end{eqnarray}
with the universal coefficients
\begin{equation}
Q^{-}c_{0}^{-}\simeq -0.7, Q^{+}c_{0}^{+}\simeq 0.9,\: Q^{\pm}c_{\infty }^{\pm }\simeq 0.5, c_{1}\simeq 1.76.
\label{A.9}
\end{equation}

To relate the magnetization to the peak structure in the specific heat we invoke Maxwell's relation
\begin{equation}
\left. \frac{\partial \left( C/T\right) }{\partial H_{c}}\right\vert
_{T}=\left. \frac{\partial ^{2}M}{\partial T^{2}}\right\vert _{H_{c}},
\label{A.10}
\end{equation}
uncovering the melting transition in terms of a singularity, while the magnetic field induced finite size effect leads to a dip. These features differ drastically from the nearly smooth behavior of the magnetization. Together with the scaling form of the specific heat (Eq.(\ref{A.6})), extended to the presence of a magnetic field,
\begin{equation}
c=\frac{A^{-}}{\alpha }\left\vert t\right\vert ^{-\alpha }f^{\pm }\left(
x\right),\: x=\frac{t}{H^{1/2\nu }},  \label{A.11}
\end{equation}
where
\begin{equation}
f_{c}^{-}\left( x\right) =\left.
\begin{array}{c}
1 \\
f_{0}^{-}\left\vert x\right\vert ^{\alpha }
\end{array}%
\begin{array}{c}
:x\rightarrow -\infty \\
:x\rightarrow 0^{-}
\end{array}\right\}.
\label{A.12}
\end{equation}
This yields the scaling form
\begin{equation}
TH_{c}^{1+\alpha /2\nu }\frac{\partial \left( c/T\right) }{\partial H_{c}}=-\frac{k_{B}A^{-}}{2\alpha \nu }x^{1-\alpha }\frac{\partial f}{\partial x}=TH_{c}^{1+\alpha /2\nu }\frac{\partial ^{2}m}{\partial T^{2}}.
\label{A.13}
\end{equation}
According to this, data plotted as $TH_{c}^{1+\alpha /2\nu }\partial \left(
c/T\right) /\partial H_{c}$ or $TH_{c}^{1+\alpha /2\nu }\partial
^{2}m/\partial T^{2}$ \emph{vs}. $x=tH^{-1/2\nu }$ should collapse on a single curve as long as the magnetic field induced finite size effect sets the limiting length ($L_{H_{c}}=\left( \Phi _{0}/aH_{c}\right) ^{1/2}<L_{ab}$).
At $T_{c}$ Eq. (\ref{A.13}) reduces to
\begin{equation}
T_{c}H_{c}^{1+\alpha /2\nu }\frac{\partial \left( c/T\right) }{\partial H_{c}}=-\frac{k_{B}A^{-}f_{0}^{-}}{2\nu }=T_{c}H_{c}^{1+\alpha /2\nu }\frac{\partial ^{2}m}{\partial T^{2}}.
\label{A.14}
\end{equation}
The scaling form (\ref{A.13}) can also be derived from Eq. (\ref{A.1}) rewritten in the form
\begin{equation}
m=-\frac{Q^{-}k_{B}\gamma }{\Phi _{0}^{3/2}}TH_{c}^{1/2}f_{m}\left( x\right),
\label{A.15}
\end{equation}
yielding
\begin{equation}
\frac{\partial m}{\partial T}=-\frac{Q^{-}k_{B}\gamma }{\Phi _{0}^{3/2}}H_{c}^{1/2}\left( f_{m}\left( x\right) +\frac{T}{T_{c}}H_{c}^{-1/2\nu }\frac{\partial f_{m}\left( x\right) }{\partial x}\right),
\label{A.16}
\end{equation}
and
\begin{equation}
T_{c}H_{c}^{-\left( \nu -2\right) /2\nu }\frac{\partial ^{2}m}{\partial T^{2}
}=-\frac{Q^{-}k_{B}\gamma }{\Phi _{0}^{3/2}}\left( 2\frac{\partial
f_{m}\left( x\right) }{\partial x}H_{c}+\frac{T}{T_{c}}\frac{\partial
^{2}f_{m}\left( x\right) }{\partial x^{2}}\right) .
\label{eqA17}
\label{A.17}
\end{equation}
Noting that
\begin{equation}
-\left( \nu -2\right) /2\nu =\left( 1+\alpha /2\nu \right),
\label{A.18}
\end{equation}
it is readily seen that close to $T_{c}$ and low magnetic fields the scaling forms (\ref{A.13}) and (\ref{A.17}) agree.

In this context it is instructive to sketch the predictions of the approximation where Gaussian fluctuations are taken into account while the magnetic field induced finite size effect is neglected. In this case the magnetization adopts the scaling form \cite{book,prange}
\begin{equation}
\frac{m}{TH^{1/2}}=-\frac{2\pi ^{1/2}k_{B}}{\Phi _{0}^{3/2}}\widetilde{f}\left( \widetilde{x}\right),
\label{A.19}
\end{equation}
where
\begin{equation}
\widetilde{x}=\frac{\Phi _{0}}{4\pi \xi _{0}^{2}H}t.  \label{A.20}
\end{equation}
Close to $\widetilde{x}=-0.5$ the scaling function adopts the form $\widetilde{f}\left( \widetilde{x}\right) =1/\left( 4\left( \widetilde{x}
+0.5\right) ^{1/2}\right) $. The resulting singularity suggests a continuous phase transition at the so-called upper critical field $H_{c2}(T)=\Phi_{0}/\left( 2\pi \xi _{0}^{2}\right) \left\vert t\right\vert $ and implies a divergence of $\partial ^{2}m/dT^{2}$ at $T_{c2}\left( H\right) $. However, taking the magnetic field induced finite size effect into account, the growth of the correlation length $\xi $ is limited by $L=\left( \Phi_{0}/\left( aH\right) \right) ^{1/2}$ and the singularity is smeared out in terms of a dip.

Furthermore, from Eq. (\ref{A.11}) we obtain for $dc/dT$ the scaling form
\begin{equation}
\frac{dc}{dT}=\frac{A^{-}}{T_{c}}H_{c}^{-\left( 1+\alpha \right) /2\nu
}\left( x^{-\left( 1+\alpha \right) /2\nu }f\left( x\right) -\alpha
x^{-\alpha }\frac{df}{dx}\right) ,  \label{A.21}
\end{equation}
whereby the data $dc/dT$ for different fields should collapse on a single curve when plotted as $H_{c}^{\left( 1+\alpha \right) /2\nu }T_{c}dc/dT$ \emph{vs}. $tH_{c}^{-1/2\nu }$ as long as $L_{H_{c}}=\left( \Phi _{0}/aH_{c}\right)^{1/2}<L_{ab}$.
To explore the the structure of this plot we consider again the Gaussian approximation. In this case the free energy density (Eq. (\ref{eq3})) adopts the form \cite{book,prange}
\begin{equation}
f_{s}=-\frac{k_{B}T}{6\pi \xi ^{3}}\widetilde{G}\left( \widetilde{z}\right),\:
\widetilde{z}=\frac{4\pi \xi _{0}^{2}H}{\Phi _{0}}=\frac{1}{\widetilde{x}},
\label{A.22}
\end{equation}
Close to $\widetilde{z}=-2$ the scaling function is given by $\widetilde{G}\left( z\right) =-\left( 1+\widetilde{z}/4\right) \left( \widetilde{z}-2\right) +const$. The resulting singularity in the specific heat, $c=-T\partial f_{s}^{2}/\partial T^{2}$, suggests again a continuous phase transition at the so-called upper critical field $H_{c2}(T)=\Phi _{0}/\left(2\pi \xi _{0}^{2}\right) \left\vert t\right\vert $ and implies a divergence of $dc/dT$ at $T_{c2}\left( H\right) $. However, taking the magnetic field induced finite size effect into account, the growth of the correlation length $\xi $ is limited by $L=\left( \Phi _{0}/\left( aH\right) \right)^{1/2}$ and the singularity is smeared out in terms of a dip.

\end{document}